\overfullrule=0pt
\input harvmac
\input slashed.sty
\def\a{{\alpha}}

\def\l{{\lambda}}
\def\lb{{\overline\lambda}}
\def\wb{{\overline w}}
\def\b{{\beta}}

\def\g{{\gamma}}

\def\d{{\delta}}
\def\e{{\epsilon}}
\def\s{{\sigma}}
\def\k{{\kappa}}

\def\half{{1\over 2}}
\def\p{{\partial}}

\def\t{{\theta}}

\def\bar{\overline}
\def\({\left(}
\def\){\right)}
\def\zd{ d^{(0)} }
\def\cF{{\cal F}}


\Title{\vbox{\hbox{AEI-2009-013 }}} {\vbox{
    \centerline{\bf The One-loop Open Superstring Massless Five-point Amplitude }
    \centerline{\bf with the Non-Minimal Pure Spinor Formalism }
 }
}

\bigskip\centerline{Carlos R. Mafra\foot{email: crmafra@aei.mpg.de} and Christian Stahn\foot{email: stahn@pas.rochester.edu}}

\bigskip

\centerline{\it ${}^1$Max-Planck-Institut f\"ur
Gravitationsphysik, Albert-Einstein-Institut}
\smallskip
\centerline{\it 14476 Golm, Germany}
\medskip
\centerline{\it ${}^2$Department of Physics and Astronomy,
University of Rochester}
\smallskip
\centerline{\it Rochester, NY 14627-0171, USA}
\medskip

\vskip .3in

We compute the massless five-point amplitude of open superstrings
using the non-minimal pure spinor formalism and obtain a simple
kinematic factor in pure spinor superspace, which can be viewed as
the natural extension of the kinematic factor of the massless
four-point amplitude.   It encodes bosonic and fermionic external
states in supersymmetric form and reduces to existing bosonic
amplitudes when expanded in components, therefore proving their
equivalence.  We also show how to compute the kinematic structures
involving fermionic states.

\vskip .3in

\Date {February 2009}

\lref\BarreiroHV{
  L.~A.~Barreiro and R.~Medina,
  ``5-field terms in the open superstring effective action,''
  JHEP {\bf 0503}, 055 (2005)
  [arXiv:hep-th/0503182].
}

\newsec{Introduction}

\lref\grosssloan{
  D.~J.~Gross and J.~H.~Sloan,
  ``The Quartic Effective Action for the Heterotic String,''
  Nucl.\ Phys.\  B {\bf 291}, 41 (1987).
}
\lref\grosswitten{
  D.~J.~Gross and E.~Witten,
  ``Superstring Modifications Of Einstein's Equations,''
  Nucl.\ Phys.\  B {\bf 277}, 1 (1986).
}

As emphasized by Gross and Sloan \grosssloan, superstring
scattering amplitudes are useful to derive the low energy
effective action of its massless modes. For example, it allows one
to obtain stringy corrections to the Yang-Mills and
Einstein-Hilbert actions \grosswitten.

Up to the year 2000 these computations were done using either the
Ramond-Neveu-Schwarz (RNS) \ref\RNS{
    P.~Ramond,
    ``Dual Theory for Free Fermions,''
    Phys.\ Rev.\  D {\bf 3}, 2415 (1971)
    \semi
    A.~Neveu and J.~H.~Schwarz,
    ``Factorizable dual model of pions,''
    Nucl.\ Phys.\  B {\bf 31} (1971) 86
    \semi
    A.~Neveu and J.~H.~Schwarz,
    ``Quark Model of Dual Pions,''
    Phys.\ Rev.\  D {\bf 4}, 1109 (1971).
} or Green-Schwarz (GS)  \ref\GS{
    M.~B.~Green and J.~H.~Schwarz,
    ``Covariant Description Of Superstrings,''
    Phys.\ Lett.\  B {\bf 136}, 367 (1984)
    \semi
     M.~B.~Green and J.~H.~Schwarz,
    ``Supersymmetrical String Theories,''
    Phys.\ Lett.\  B {\bf 109}, 444 (1982).
} formulation in ten-dimensional flat background.  In the open
(closed) superstring, the tree and one-loop four-gluon
(four-graviton) amplitudes lead to an infinite series of higher
derivative corrections starting with the well-known $t_8F^4$
($t_8t_8R^4$) interaction terms \grosswitten\ref\schwarz{
  J.~H.~Schwarz,
  ``Superstring Theory,''
  Phys.\ Rept.\  {\bf 89}, 223 (1982).
}. After several years of efforts, the corresponding corrections
were finally computed at the two-loop level in a series of papers
by D'Hoker and Phong (see \ref\dhoker{
    E.~D'Hoker and D.~H.~Phong,
    ``Two-Loop Superstrings VI: Non-Renormalization Theorems and the 4-Point
    Function,''
    Nucl.\ Phys.\  B {\bf 715}, 3 (2005)
    [arXiv:hep-th/0501197]
} and references therein), starting with terms of the form $t_8
D^2F^4$ ($t_8t_8 D^4 R^4$).

The results above were obtained for the bosonic amplitudes in the
NS and NS-NS sectors. While the fermionic four-point amplitudes
involving strings in the R sector are known at tree- and one-loop
levels, the two-loop computations using the gauge slice
independent methods of \dhoker\ are still lacking in the RNS
formalism.  The problems involving fermionic amplitudes present a
major complication to the supersymmetric completion to string
effective actions and the analysis of higher-order derivative
corrections for RR gauge fields (see e.g. \ref\PeetersUB{
  K.~Peeters, P.~Vanhove and A.~Westerberg,
  ``Chiral splitting and world-sheet gravitinos in higher-derivative string
  amplitudes,''
  Class.\ Quant.\ Grav.\  {\bf 19}, 2699 (2002)
  [arXiv:hep-th/0112157]
\semi
  K.~Peeters, P.~Vanhove and A.~Westerberg,
  ``Supersymmetric higher-derivative actions in ten and eleven dimensions,  the
  associated superalgebras and their formulation in superspace,''
  Class.\ Quant.\ Grav.\  {\bf 18}, 843 (2001)
  [arXiv:hep-th/0010167].
}).

This state of affairs, however, started to change with the
proposal of the pure spinor formalism \ref\superp{
  N.~Berkovits,
  ``Super-Poincare covariant quantization of the superstring,''
  JHEP {\bf 0004}, 018 (2000)
  [arXiv:hep-th/0001035].
}. It does not suffer from the inherent difficulties of the RNS
and GS formulations responsible for the {\it status quo}
concerning two-loop fermionic amplitudes.  It accomplishes that by
having manifest space-time supersymmetry together with a Lorentz
invariant BRST quantization procedure\foot{The usefulness of the
PS formalism is of course not restricted to amplitude
calculations, see e.g. the review \ref\ictp{
  N.~Berkovits,
  ``ICTP lectures on covariant quantization of the superstring,''
  [arXiv:hep-th/0209059].
}.}.

It was soon realized that tree-level amplitudes computed using this
new formalism were equivalent to RNS results \ref\brennotree{
  N.~Berkovits and B.~C.~Vallilo,
  ``Consistency of super-Poincare covariant superstring tree amplitudes,''
  JHEP {\bf 0007}, 015 (2000)
  [arXiv:hep-th/0004171].
}. When the multiloop prescription came into existence 
four years later, it was readily used to obtain a non-renormalization theorem\foot{This
theorem was later extended in \ref\nonrenor{
  N.~Berkovits,
  ``New higher-derivative R**4 theorems,''
  Phys.\ Rev.\ Lett.\  {\bf 98}, 211601 (2007)
  [arXiv:hep-th/0609006].
} (see also \ref\basu{
 A.~Basu,
  ``The D**10 R**4 term in type IIB string theory,''
  Phys.\ Lett.\  B {\bf 648}, 378 (2007)
  [arXiv:hep-th/0610335].
}) and used by Green, Russo and Vanhove \ref\GRV{
 M.~B.~Green, J.~G.~Russo and P.~Vanhove,
  ``Ultraviolet properties of maximal supergravity,''
  Phys.\ Rev.\ Lett.\  {\bf 98}, 131602 (2007)
  [arXiv:hep-th/0611273].
} to argue that $N = 8$ SUGRA  is finite up to 8 loops.} regarding
the $R^4$ term in the effective action \ref\multiloop{
  N.~Berkovits,
  ``Multiloop amplitudes and vanishing theorems using the pure spinor
  formalism for the superstring,''
  JHEP {\bf 0409}, 047 (2004)
  [arXiv:hep-th/0406055].
} and the two-loop massless four-point amplitude~\ref\twoloop{
  N.~Berkovits,
  ``Super-Poincare covariant two-loop superstring amplitudes,''
  JHEP {\bf 0601}, 005 (2006)
  [arXiv:hep-th/0503197].
} of the type IIB superstring.  The two-loop kinematic factor
derived in \twoloop\ was expressed as an integral over pure spinor
superspace \ref\explsuper{
  N.~Berkovits,
  ``Explaining pure spinor superspace,''
  [arXiv:hep-th/0612021].
} and was subsequently shown \ref\PSequivII{
    N.~Berkovits and C.~R.~Mafra,
        ``Equivalence of two-loop superstring amplitudes in the pure spinor and RNS
        formalisms,''
        Phys.\ Rev.\ Lett.\  {\bf 96}, 011602 (2006)
        [arXiv:hep-th/0509234].
} to agree with the two-loop RNS computations of D'Hoker and Phong
\dhoker.  Furthermore, explicit bosonic computations also proved
the PS/RNS equivalence at one loop \ref\oneloop{
  C.~R.~Mafra,
  ``Four-point one-loop amplitude computation in the pure spinor formalism,''
  JHEP {\bf 0601}, 075 (2006)
  [arXiv:hep-th/0512052].
} and were followed by calculations involving fermionic external
states in \ref\stahn{
  C.~Stahn,
  ``Fermionic superstring loop amplitudes in the pure spinor formalism,''
  JHEP {\bf 0705}, 034 (2007)
  [arXiv:0704.0015 [hep-th]].
}.

In 2008, several manipulations in pure spinor superspace were used
to prove that the massless four-point amplitudes at tree-level and
at one and two loops are all related to each other \ref\mafraids{
  C.~R.~Mafra,
  ``Pure Spinor Superspace Identities for Massless Four-point Kinematic
  Factors,''
  JHEP {\bf 0804}, 093 (2008)
  [arXiv:0801.0580 [hep-th]].
}, being proportional (up to Mandelstam invariants) to the {\it
same} tree-level kinematic factor. With the identities of
\mafraids, the knowledge of all bosonic and fermionic four-point
scattering amplitudes up to two loops was obtained altogether
simply by possessing the complete tree-level result.

By the end of 2005, Berkovits proposed an extension of the pure
spinor formalism by introducing non-minimal variables, which
resulted in a simplified expression for the b-ghost~\ref\NMPS{
    N.~Berkovits,
        ``Pure spinor formalism as an N = 2 topological string,''
    JHEP {\bf 0510}, 089 (2005)
    [arXiv:hep-th/0509120].
}. In this non-minimal formalism, the b-ghost is one of the
generators of a ${\hat c} =3$ $N=2$ topological algebra, and it is
``simpler'' than the corresponding b-ghost of the minimal
formalism, albeit still ``complicated'' due to its
composite nature, with poles in~$1/(\l\lb)$.
\lref\anom{
  N.~Berkovits and C.~R.~Mafra,
  ``Some superstring amplitude computations with the non-minimal pure spinor
  formalism,''
  JHEP {\bf 0611}, 079 (2006)
  [arXiv:hep-th/0607187].
} \lref\sken{
  J.~Hoogeveen and K.~Skenderis,
  ``BRST quantization of the pure spinor superstring,''
  JHEP {\bf 0711}, 081 (2007)
  [arXiv:0710.2598 [hep-th]].
} 
It was already argued in \NMPS\
that multiloop amplitudes computed using the new topological
methods should be equivalent to the results computed in the minimal
formalism, and that was verified by some explicit amplitude 
computations in \anom\foot{
There exists now a formal proof of NMPS/PS
equivalence by Hoogeveen and Skenderis \sken.
}.

\lref\nekr{
  N.~Berkovits and N.~Nekrasov,
  ``Multiloop superstring amplitudes from non-minimal pure spinor formalism,''
  JHEP {\bf 0612}, 029 (2006)
  [arXiv:hep-th/0609012].
}\lref\Rpuri{
    G.~Policastro and D.~Tsimpis,
        ``R**4, purified,''
        [arXiv:hep-th/0603165].
}

With the non-minimal pure spinor framework being simpler than its
version of five years earlier and having a prescription which is
in principle valid to all-loop orders \nekr, a renewed computation
of superstring scattering amplitudes is a task worth pursuing and
holds the potential to increase our understanding of stringy
corrections to low-energy effective actions (see e.g. the
treatment of tree-level four-point amplitudes in \Rpuri).

{}From the inner perspective of the non-minimal pure spinor
formalism, the straightforward way with which the expressions for
the massless four-point amplitudes were obtained (up to two loops)
is related to the fact that in these cases only the zero modes of
the b-ghost played a r\^ole.  This allowed the amplitudes in pure
spinor superspace to be determined essentially by powerful
symmetry arguments \multiloop\twoloop\anom.

The motivation which started this paper was to test the abilities
of the pure spinor formalism in a situation where not only the
b-ghost zero-modes would (in principle) contribute. In view of the
complicated nature of the b-ghost, that motivation was deemed to
be another non-trivial test whose result was worth knowing. At the
same time, the knowledge about higher-point amplitudes involving
fermionic external states would benefit in case of a successful
outcome.
As we will show, the (non-minimal) pure spinor formalism\foot{ In
the this paper we will simply refer to the non-minimal pure spinor
formalism as ``pure spinor formalism'' (PS).} withstands the test
and leads to a simple answer in pure spinor superspace.

This paper is organized as follows. In section 2 we recall the
prescription of the PS formalism for the computation of the open
superstring one-loop amplitude. We propose simplified expressions
for the pure spinor integration measures and the normalization
factor ${\cal N}(y)$ to facilitate explicit computations, allowing
one to identify a shortcut which can be taken to integrate out the
$w_\a, \wb^\a$ and $s^\a$ variables. In section 3 we prove the
vanishing of the contributions coming from the non zero-modes of
the b-ghosts and obtain the kinematic factor of this five-point
amplitude as a (rather large) expression in pure spinor
superspace, relegating the detailed calculations to Appendix B. We
then prove its gauge invariance by showing that the gauge
variation is a total derivative which vanishes by the cancelled
propagator argument. In section 4 we simplify the answer of
section 3 by using further pure spinor manipulations to the point
of being able to identify a simple form for the massless
five-point kinematic factor,
$$
K_5 = \sum_{j=2}^5 L_{1j}\eta(z_1,z_j) + \sum_{2\le i<j\le 5}
K_{ij}\eta(z_i,z_j)
$$
where
  $$\eqalign{
  L_{12} &= - \Bigl\langle
  \Big[(\l A^1)(k^1\cdot A^2) + A^1_p (\l\g^p W^2) \Big]
  (\l\g^m W^5)(\l\g^n W^3){\cal F}^4_{mn}
  \Bigr\rangle \cr
  K_{25} &= - \Bigl\langle
  (\l A^1) \Big[
  (\l\g^m W^2)(k^2\cdot A^5) - {1\over 4} (\l\g^m\g^{rs} W^5) \cF^2_{rs} \Big]
  (\l\g^n W^3){\cal F}^4_{mn} \Bigr\rangle - (2\leftrightarrow 5)
  }$$
and similarly for other choices of labels.  The above two
expressions can be understood as coming from the OPE computation
of the natural generalization
  \eqn\naturalis{
  \bigl\langle (\l A^1)(\l\g^m W^2)(\l \g^n W^3){\cal F}^4_{mn}\, U^5
  \bigr\rangle
  }
of the massless four-point kinematic factor $\langle (\l
A^1)(\l\g^m W^2)(\l \g^n W^3){\cal F}^4_{mn} \rangle$.

In section 5 we calculate the (bosonic) component expansion of the
``kinematic factor''~\naturalis\ and show that it agrees, up to
total derivative terms, with previous (bosonic\foot{We also
acknowledge the existence of fermionic results obtained by Lin
{\it et al.~}\ref\lin{
  Z.~H.~Lin,
  ``One Loop Closed String Five Particle Fermion Amplitudes in the Covariant
  Formulation,''
  Int.\ J.\ Mod.\ Phys.\  A {\bf 5}, 299 (1990)
  \semi
  Z.~H.~Lin, L.~Clavelli and S.~T.~Jones,
  ``Five Point Function in the Covariant Formulation of the Type I Superstring
  Theory,''
  Nucl.\ Phys.\  B {\bf 294}, 83 (1987).
} using the methods of Atick and Sen \ref\sen{
  J.~J.~Atick and A.~Sen,
  ``Covariant one loop fermion emission amplitudes in closed string theories,''
  Nucl.\ Phys.\  B {\bf 293}, 317 (1987).
}, but we have not tried to bring their results to a form which
could be readily compared to the fermionic results obtained from
\naturalis. As the pure spinor formalism is manifestly
supersymmetric, we are guaranteed that once the bosonic amplitudes
are verified to be correct then so are the fermionic ones.})
five-point results \ref\leeSiegel{
   K.~Lee and W.~Siegel,
   ``Simpler superstring scattering,''
   JHEP {\bf 0606}, 046 (2006)
   [arXiv:hep-th/0603218].
}\ref\richards{
  D.~M.~Richards,
  ``The One-Loop Five-Graviton Amplitude and the Effective Action,''
  JHEP {\bf 0810}, 042 (2008)
  [arXiv:0807.2421 [hep-th]].
}\ref\frampton{
  P.~H.~Frampton, P.~Moxhay and Y.~J.~Ng,
  ``Explicit Evaluation Of Pentagon Diagram For Open Superstrings,''
  Nucl.\ Phys.\  B {\bf 276}, 599 (1986).
}\ref\tsuchiya{
 A.~Tsuchiya,
  ``More on One Loop Massless Amplitudes of Superstring Theories,''
  Phys.\ Rev.\  D {\bf 39}, 1626 (1989).
}. We also compute some two- and four-fermion terms.

In Appendix A we review\foot{This covariant proof was
first obtained in \ref\tese{
    C.~R.~Mafra,
    ``Superstring Scattering Amplitudes with the Pure Spinor Formalism'', Ph.D thesis,
    September 2008, arXiv:0902.1552 [hep-th].
}.} the covariant (i.e. without relying on a U(5) decomposition as
in \anom) proof of equivalence between the minimal and non-minimal
one-loop massless four-point amplitude, as similar manipulations
will be used again in the five-point kinematic factor computation of
Appendix B. In Appendix C -- for completeness -- we show how to
covariantly compute expressions of the form $\langle \l^4\lb
\t^5\rangle$ and in Appendix D we present the technicalities
needed when computing component expansions of pure spinor
superspace expressions. In particular, the kinematic reduction
algorithms of \stahn\ are expanded to deal with the structures
appearing at five-points. They are well-suited for implementations
using a computer algebra system such as FORM \ref\FORM{
  J.~A.~M.~Vermaseren,
  ``New features of FORM,''
  [arXiv:math-ph/0010025]
 \semi
  M.~Tentyukov and J.~A.~M.~Vermaseren,
  ``The multithreaded version of FORM,''
  [arXiv:hep-ph/0702279].
}\ref\pss{
C.~R.~Mafra, ``PSS: A FORM Program to Compute Pure
Spinor Superspace Expressions'', unpublished.
}
and/or Mathematica (using Ulf Gran's GAMMA package \ref\GAMMA{
  U.~Gran,
  ``GAMMA: A Mathematica package for performing Gamma-matrix algebra and  Fierz
  transformations in arbitrary dimensions,''
  [arXiv:hep-th/0105086].
}).

\newsec{The scattering amplitude prescription}

The non-minimal pure spinor formalism prescription for the
massless five-point open superstring amplitude at one-loop is
given by \NMPS\ \eqn\fivept{ {\cal A} = \sum_{\rm top}\int dt
\Bigl\langle {\cal N}(y) \Bigl(\int d^2w \mu(w) b(w)\Bigr)
V^1(0)\prod_{I=2}^5 \int dz_I U^I(z_I) \Bigr\rangle. } Here
$V^1(z) = \l^\a A^1_\a(x,\t)$ denotes the unintegrated vertex
operator and
  \eqn\integrado{ U(z) = \p\t^{\a}A_{\a}(x,\t) +
  A_m(x,\t)\Pi^m + d_{\a}W^{\a}(x,\t) + \half N_{mn}{\cal
  F}^{mn}(x,\t)
  }
the integrated vertex operator, $\mu(w)$ is the Beltrami
differential with conformal weight $(1,-1)$ and the sum is over
the three Riemann surfaces which describe the one-loop interaction
of open strings: the planar and non-planar cylinders and the
M\"obius strip. The b-ghost $b(w)$ is a composite field whose
expression reads \NMPS, \eqn\bghost{ b = s^{\a}\p\lb_{\a} +
{2\Pi^m(\lb\g_m d)-N_{mn}(\lb\g^{mn}\p\t)-J(\lb\p\t) -(\lb\p^2\t)
\over 4(\lb\l)} }
$$
+{ (\lb\g^{mnp}r)(d\g_{mnp}d+24N_{mn}\Pi_p) \over 192(\lb\l)^2}
-{(r\g^{mnp}r)(\lb\g_m d)N_{np} \over 16(\lb\l)^3}
+{(r\g^{mnp}r)(\lb\g_{pqr}r)N_{mn}N^{qr} \over 128(\lb\l)^4}.
$$
The factor ${\cal N}(y)$ is needed to regularize the integration
over the non-compact domain of the pure spinors, and we will use
  \eqn\newN{
  {\cal N} = \exp\big[-(\l\lb) - (r\t) - (w
  {\bar w}) + (s d)\big]
  = \exp(\{Q,\chi\})
  \, \hbox{ for } \, \chi = -(\lb\t) - (w s)
  }
This regularization is different from the one presented in \NMPS\
because our choice of $\chi$ is not gauge invariant. However, as
it is still valid that ${\cal N}(y) = 1 + Q\Omega$ for some
$\Omega$, the integral will be independent of the choice for
$\chi$.

The evaluation of \fivept\ will give rise to an expression of the
form
$$
{\cal A} = \int [d\l][d\lb][dr] [dw][d\wb][ds](d^{16}d)d^{16}\t
\,{\cal N} f(r,\t),
$$
where the non-zero modes have already been integrated out through
their OPEs. The zero modes are to be integrated with the following
measures\foot{They are written slightly differently from the
original expressions of \NMPS\ (see e.g. \tese)}:
\eqnn\dl \eqnn\dlb \eqnn\dr
\eqnn\dw \eqnn\dwb \eqnn\ds
$$
\eqalignno{
[d\l] & = (\l\lb)^{-3} (\e_{16}\cdot d^{11}\l)_{\k_1{\ldots} \k_5}
(\lb\g^m)^{\k_1}(\lb\g^n)^{\k_2}(\lb\g^p)^{\k_3}
(\g_{mnp})^{\k_4\k_5} & \dl \cr
[d\lb] & = (\l\lb)^{-3} (\e_{16}\cdot d^{11}\lb)^{\k_1{\ldots} \k_5}
(\l\g^m)_{\k_1}(\l\g^n)_{\k_2}(\l\g^p)_{\k_3}
(\g_{mnp})_{\k_4\k_5} & \dlb \cr
[dr] & = \e_{\a_1{\ldots} \a_{11}\k_1{\ldots} \k_5} (\lb\g^m)^{\k_1}
(\lb\g^n)^{\k_2}(\lb\g^p)^{\k_3} \g_{mnp}^{\k_4\k_5}\, \p^{\a_1}_r {\ldots}  \p^{\a_{11}}_r & \dr\cr
[dw] & = (\l\g^m)_{\k_1}(\l\g^n)_{\k_2}(\l\g^p)_{\k_3}(\g_{mnp})_{\k_4\k_5}\,
\e^{\k_1{\ldots} \k_5\rho_1{\ldots} \rho_{11}} dw_{\rho_1}
{\ldots}  dw_{\rho_{11}} & \dw \cr
[d\wb] & =
(\lb\g^m)^{\k_1}(\lb\g^n)^{\k_2}(\lb\g^p)^{\k_3}
(\g_{mnp})^{\k_4\k_5}\, \e_{\k_1{\ldots} \k_5\a_1{\ldots} \a_{11}}
d\wb^{\a_1}{\ldots}  d\wb^{\a_{11}} & \dwb \cr
[ds] & =
(\l\lb)^{-3}(\l\g^m)_{\k_1}(\l\g^n)_{\k_2}(\l\g^p)_{\k_3}(\g_{mnp})_{\k_4\k_5}
\e^{\k_1{\ldots} \k_5\rho_1{\ldots} \rho_{11}} \p^s_{\rho_1}
{\ldots}  \p^s_{\rho_{11}} & \ds
}
$$
Note that the measure \dw\ is
gauge invariant under $\d w_{\a} = (\l\g^m)_{\a} \Omega_m$ due to
$ (d\l\g^q)_{[\d_1}(\l\g^m)_{\k_1}(\l\g^n)_{\k_2}(\l\g^p)_{\k_3}
(\g_{mnp})_{\k_4\k_5]} =0, $ because there is no vector
representation in the decomposition of $\l^4\t^6$.

Apart from an overall coefficient which is not determined, one can
use a shortcut to perform the integrations over $[dw], [d\wb]$ and
$[ds]$. At one loop there are eleven zero modes of $s^\a$, which
can only\foot{The term $s^\a\p\lb_\a$ of the b-ghost does not
contribute because there is no $\wb^\a$ in the external vertices
to kill the non zero-modes of $\p\lb_\a$.} come from ten factors
of the term $(s^{(0)} \zd)$ in the regulator ${\cal N}$. Therefore
the remaining five $d_\a$ zero modes must come from the b-ghost
and the external vertices. By ghost number conservation we obtain
  $$\displaylines{
  \quad \int d^{16}d[dw][d\wb][ds] {\rm e}^{-(w\wb) + (sd) - (\l\lb) -
  (r\t)} \zd_{\k_1}{\ldots}  \zd_{\k_5} f^{\k_1{\ldots}
  \k_5}(r,\t)
  = (\l^3)_{[\k_1{\ldots}\k_5]} f^{\k_1{\ldots} \k_5}(r,\t)
  }$$
where $(\l^3)_{[\k_1\k_2\k_3\k_4 \k_5]}$ is some tensor with five
antisymmetric free indices containing three pure spinors. The
unique such tensor is proportional to
$(\l\g_m)_{\k_1}(\l\g_n)_{\k_2}(\l\g_p)_{\k_3}(\g^{mnp})_{\k_4\k_5}$.
One can thus see that the net effect of evaluating the pure spinor
measures $[dw], [d\wb]$, $[ds]$ and $d^{16}d$ is to replace the five $\zd_\a$
zero-modes from the b-ghost and external vertices by \eqn\truque{
\zd_{\k_1}\zd_{\k_2}\zd_{\k_3}\zd_{\k_4}\zd_{\k_5} \longrightarrow
(\l\g_m)_{\k_1}(\l\g_n)_{\k_2}(\l\g_p)_{\k_3}(\g^{mnp})_{\k_4\k_5}.
}

\newsec{The massless five-point amplitude computation}

\subsec{Zero mode saturation and OPEs}

The five $d_{\a}$ zero modes from the b-ghost and the external
vertices appearing in \truque\ can be obtained by four distinct
ways.  Each one of them will imply a different ``kinematic
factor'',
  \eqnn\If\
  $$\eqalignno{
  I_1 & = {1\over 2}\Bigl\langle {\Pi^m(z_0) \over (\l\lb)}(\lb \g_m \zd)
       (\l A^1)(\zd W^2)(\zd W^3) (\zd W^4)(\zd W^5) \Bigr\rangle \cr
  I_2 & = -{1\over 16}\Bigl\langle {(r\g_{mnp}r) \over(\l\lb)^3}
        (\lb\g^m \zd)N^{np}(z_0) (\l A^1)(\zd W^2)(\zd W^3)
        (\zd W^4)(\zd W^5) \Bigr\rangle \cr
  I_3 & = {1\over 96}\Bigl\langle {(\lb\g_{mnp}r) \over(\l\lb)^2}(\zd\g^{mnp}
        {\hat d}(z_0)) (\l A^1)(\zd W^2)(\zd W^3)(\zd W^4)(\zd W^5) \Bigr\rangle\cr
  I_4 & = {1 \over 192}\Bigl\langle{(\lb\g_{mnp}r)\over(\l\lb)^2}
         (\zd\g^{mnp}\zd) (\l A^1)(\zd W^2)(\zd W^3)(\zd W^4) \cr
  & \quad\quad\quad\times \bigl(A^5_q\Pi^q + ({\hat d}W^5)
  + {1\over 2}N^{mn}{\cal F}^5_{mn}\bigr) \Bigr\rangle + \hbox{cycl(2345)}, & \If
  }
  $$
where we have written $d_\a(z) = d^{(0)}_\a \omega(z) + {\hat
d}_\a(z)$, with $\omega(z)$ being the holomorphic one-form and
$d^{(0)}_\a$ the 16 zero modes of $d_\a$. In the expressions above
one has to integrate out the conformal weight one variables
through their OPEs and use the measures \dl\ -- \ds\ to deal with
the remaining zero modes, but we have omitted them as it is clear
from the context that they are there\foot{We also don't write the
integration over the vertex positions in most of our formulae to
avoid cluttering. One should note that the expressions we call
``kinematic factors'' also depend on the coordinates $z_j$ of the
vertices, so that is a slight abuse of terminology.}.

As discussed in \ref\verlindes{
  E.~P.~Verlinde and H.~L.~Verlinde,
  ``Chiral bosonization, determinants and the string partition function,''
  Nucl.\ Phys.\  B {\bf 288}, 357 (1987).
}, if the scalar propagator is $f(z,w) = \langle X(z,{\bar
z})X(w,{\bar w})\rangle$ then the OPE of a $(1,0)$-system with the
zero modes projected out is given by $-\p_z f(z,w)$. Furthermore
one can show that up to terms which drop out of correlation
functions \ref\polchinski{
  J.~Polchinski,
  ``String theory. Vol. 1: An Introduction to the Bosonic String,''
{\it  Cambridge, UK: Univ. Pr. (1998) 402 p} } the function
$f(z,w)$ in the torus is given in terms of the prime form $E(z,w)$
by \eqn\logchi{ f(z,w) = - \ln(|E(z,w)|^2) + 2\pi { ({\rm
Im}(z-w))^2\over {\rm Im}\tau}, } whereas for the cylinder and
M\"obius strip it is obtained through proper identifications under
involutions \polchinski. We also define the $\eta(z,w)$ function
by \eqn\opes{ \langle {\hat d}(z)\t(w)\rangle = -\langle \Pi
(z)X(w,{\bar w})\rangle = - {\p\over \p z} f(z,w) \equiv
\eta(z,w), } and the explicit expansions of both $E(z,w)$ and
$\eta(z,w)$ in terms of Jacobi theta functions will not be needed
here but they can be deduced from the formulae in \ref\gswII{
  M.~B.~Green, J.~H.~Schwarz and E.~Witten,
  ``Superstring Theory. Vol. 2: Loop Amplitudes, Anomalies And Phenomenology,''
{Cambridge, UK: Univ. Pr. (1987) 596 p. (Cambridge Monographs On
Mathematical Physics)} }.

\subsec{The vanishing of b-ghost OPEs}

We begin by noticing that after the OPEs are computed in $I_1,
I_2$ and $I_3$ the result is actually a total derivative in $w$,
  \eqnn\zerod
  $$\eqalignno{
  I_1 + I_2 + I_3 &\propto
  \int d^2w \sum_{j=2}^5 K_{0j} \eta(w,z_j)
  \exp{\bigl[\sum_{i<j}^5 (k^i\cdot k^j)f(z_i,z_j)\bigr]} \cr
  &\propto \int d^2w {\p\over \p w}\Bigl[ \sum_{j=2}^5 K_{0j}
  f(w,z_j) \exp{\bigl[\sum_{i<j}^5 (k^i\cdot k^j)f(z_i,z_j)\bigr]}\Bigr]
  \,, & \zerod
  }$$
for some kinematic factors $K_{0j}$, where $w$ and $z_j$ are the
positions of the b-ghost and the external vertices, respectively.
In \zerod\ we used \opes\ and reinstated the integral over $w$
because it will be needed to prove that it vanishes for the
topologies considered in \fivept. At this point it is instructive
to see what one would obtain if the Riemann surface in \zerod\
were the torus instead of, say, the cylinder. In this case the
integration in \zerod\ clearly vanishes because there can't be any
contribution from non-trivial cycles, as the function $f(w,z_j)$
is doubly periodic over the torus. Now to see the vanishing of
\zerod\ for the cylinder one has to notice\foot{CRM would like to
thank Nathan Berkovits for discussions on this point.} that the
b-ghost in $\int d^2w \, b(w)$ is already implicitly defined via
the ``doubling trick'' to live in the double of the cylinder,
which is the torus. So if the cylinder is the region \polchinski\
$$
0 \le {\rm Re}(w) \le \pi, \quad w \sim w + 2\pi i t,
$$
then both $b(w)$ and $\tilde{b}(\wb)$ are defined {\it only} in
this region. The doubling trick in this case, however, consists in
trading $b(w)$ and $\tilde{b}(\wb)$ by an augmented $b(w)$ defined
also in the reflected region $-\pi \le {\rm Re}(w) \le 0$, through
the following identification
$$
b(w) = \tilde{b}(\wb'), \quad -\pi \le {\rm Re}(w) \le 0, \quad
w'= - \wb.
$$
Now the integral of \zerod\ over this ``doubled cylinder'' results
in two contributions over the boundaries at ${\rm Re}(w) = \pm
\pi$, which cancel each other out because of the periodicity
$f(w,z_j) = f(w+2\pi,z_j)$.

\subsec{The kinematic factor in pure spinor superspace}

As shown above, there are no contributions from $I_1,{\ldots}
,I_3$ and therefore the five-point kinematic factor is obtained
entirely from $I_4$. This will be done by computing the OPEs
between the conformal weight-one variables and using \truque\ to
deal with the remaining five $d_\a$ zero-modes. Considering first
the OPEs which do not involve the (fixed) position of the
unintegrated vertex we restrict our attention to the terms
proportional to e.g., $\eta(z_2,z_5)$,
  $$\displaylines{
  \Bigl\langle {(\lb\g_{mnp}r) \over (\l\lb)^2}(d\g^{mnp}d)
  (\l A^1)(dW^2)(dW^3)(dW^4) (A^5_q\Pi^q + ({\hat d}W^5))
  \Bigr\rangle + ( 2 \leftrightarrow 5) \cr
  = \eta(z_5,z_2) L_{52} + \eta(z_2,z_5) L_{25} \equiv \eta(z_2,z_5)
  K_{25}
  }$$
where, upon restoring the suppressed notation,
  \eqnn\arr
  $$\eqalignno{
  K_{25} &= \int [d\l][d\lb][dr] {\rm e}^{-(\l\lb) - (r\t)}
  {(\lb\g_{mnp}r) \over (\l\lb)^2}\Big[ {1\over 4} (\l A^1)(\l\g^m
  \g^{rs} W^5)(\l \g^n W^3)(\l\g^p W^4){\cal F}^2_{rs} \cr
  &\qquad\qquad
  -k^2_q (\l A^1)(\l\g^m W^2)(\l\g^n W^3)(\l\g^p W^4)
  A^5_q\Big]  - (2\leftrightarrow 5) \,. & \arr\
  }$$
To arrive at \arr\ one uses \truque\ and notes that $\eta(z_2,z_5)
= - \eta(z_5,z_2)$. Finally from $r_\a \e^{-(r\t)} = -
D_\a\e^{-(r\t)}$ one can replace $r_\a$ by $D_\a$ in \arr. So we
have expressed the one-loop kinematic factor computation in terms
of a tree-level pure spinor superspace integral, which is amenable
to various simplifications through the use of pure spinor
identities. As will become clear during the computations, the
presence of $\lb_\a$ in the superspace integral~\arr\ will not
play an important role: Using sufficiently many pure spinor
superspace manipulations, they can always be seen to appear as
factors of $(\l\lb)$ in the terms which contribute to the
kinematic factor\foot{As a consequence -- once we prove that the
the $\lb_\a \l^\b$ pair appear contracted as $(\l\lb)$ --
we essentially ignore their presence in our formulae, as they
will only affect the overall coefficient.}

The task is now to compute
  \eqnn\pili
  $$\eqalignno{
  (\l\lb)^2 K_{25} &=  {1\over 4} \bigl\langle(\lb\g_{mnp}D)\bigl[
  (\l A^1){\cal F}^2_{rs} (\l\g^m \g^{rs} W^5)(\l\g^n W^3)(\l\g^p
  W^4)\bigr]\bigr\rangle & \pili\ \cr
  & \quad {}- \bigl\langle (\lb\g_{mnp}D)\bigl[
  (k^2\cdot A^5)(\l A^1) (\l\g^m W^2)(\l\g^n W^3)(\l\g^p W^4)
  \bigr]\bigr\rangle - (2\leftrightarrow 5) \,.
  }$$
As shown in Appendix B by using the action of the BRST operator
$$
Q{\cal F}_{mn} = 2k_{[m} (\l\g_{n]} W), \quad Q W^{\a} = {1\over
4}(\l\g^{mn})^{\a}{\cal F}_{mn},\quad QA_m = (\l\g_m W) + k_m(\l
A) \,,
$$
the pure spinor constraint $(\l\g^m\l) = 0$ and various gamma
matrix identities we obtain
  \eqnn\princ
  $$\eqalignno{
  (\l\lb)^2 K_{25} &= - 16\bigl\langle(\l\lb)(\l A^1)(\l\g^m W^2)(\l\g^n
  W^4){\cal F}^3_{mn} (k^2\cdot A^5)\bigr\rangle \cr
  &\quad
  + 4 \bigl\langle(\l\lb)(\l A^1)(\l\g^m \g^{rs} W^5)(\l\g^n W^4){\cal
  F}^2_{rs}{\cal F}^3_{mn}\bigr\rangle \cr
  &\quad
  - 12\bigl\langle(\l\lb)(\l
  A^1)(\l\g^m W^2)(\l\g^n W^3){\cal F}^4_{mn}(k^2\cdot A^5)\bigr\rangle
  \cr
  &\quad
  + 3 \bigl\langle(\l\lb)(\l A^1)(\l\g^m \g^{rs} W^5)(\l\g^n W^3){\cal
  F}^2_{rs}{\cal F}^4_{mn}\bigr\rangle \cr
  &\quad
  - 12\bigl\langle(\l\lb)(\l A^1)(\l\g^{m} W^3)(\l\g^{n} W^4){\cal
  F}^2_{mn}(k^2\cdot A^5)\bigr\rangle \cr
  &\quad
  - 24 k^2_m\bigl\langle(\l\lb)(\l A^1)(\l\g^{[m} W^3)(\l\g^{n]} W^4)(W^5
  \g_n W^2)\bigr\rangle \cr
  &\quad
  - 12 \bigl\langle(\l\lb)(\l A^1)(\l\g^{[m} W^3)(\l\g^{n]} W^4){\cal
  F}^2_{mt}{\cal F}^5_{nt}\bigr\rangle \cr
  &\quad
  + 2(k^2\cdot k^5)\bigl\langle(\l A^1)(\l A^5)(\l\g^m W^2)(\l\g^n
  W^4)(\lb\g_{mn}W^3)\bigr\rangle \cr
  &\quad
  + 2 (k^2\cdot k^5)\bigl\langle(\l\lb)(\l A^5)(\l\g^m W^2)(\l\g^n
  W^4)(A^1\g_{mn}W^3)\bigr\rangle \cr
  &\quad
  - 2 (k^2\cdot k^5)\bigl\langle(\l\lb)(\l A^1)(\l\g^m W^2)(\l\g^n
  W^4)(A^5\g_{mn}W^3)\bigr\rangle - (2\leftrightarrow 5) \,
  & \princ
  }$$

As mentioned before, the pure spinor $\lb_\a$ appears only in
overall factors of $(\l\lb)$ except for one term -- which will
turn out to be part of a total derivative, as also shown in
Appendix~B. However \princ\ is not a ``simple'' expression by any
standard, and it would be desirable to seek for cancellations of
terms by manipulations in pure spinor superspace. This will be
done after we prove the gauge invariance of the amplitude.

\subsec{Gauge invariance}

The one-loop scattering amplitude prescription in the non-minimal
pure spinor formalism \NMPS\ uses one non-integrated vertex
operator and therefore is not manifestly gauge invariant.
Nevertheless gauge invariance can be proved after the integration
over the vertex operator positions is performed, as one can easily
verify from the prescription. One can also see this explicitly for
the kinematic factor \princ\ as follows. From the five-point
amplitude prescription of \fivept\ it follows that under the gauge
variation of $\d A^1_\a = D_\a \Omega^1$,
  \eqn\zerog{
  \d {\cal A} = \int dt \Bigl\langle
   {\cal N} \int \left(b\cdot \mu\right)
   \Omega^1\int U^2\int U^3\int U^4\int \p(\l A^5) \Bigr\rangle
   + {\rm cycl(2345)} = 0.
  }
In the above we ``integrated the BRST charge by parts'', used
the identity $QU^5 = \p (\l A^5)$ \superp\ and the fact that the
contribution from $\{Q, b\} = T$ vanishes identically due to the
lack of $d_\a$ zero modes \anom. Furthermore, the vanishing of \zerog\
is justified by the
cancelled propagator argument. Using $\p (\l A^5) = \p \t^\a
\p_\a(\l A^5) + \Pi^m k^5_m (\l A^5)$ one gets the terms
proportional to $\eta(z_2,z_5)$,
  \eqn\totderiv{
   (k^2\cdot k^5) \, \bigl\langle (\lb \g_{mnp} D)\big[
  \Omega^1 (\l A^5)(\l\g^m W^2)(\l\g^n W^3)(\l\g^p W^4) \big] \bigr\rangle -
  (2\leftrightarrow 5)\simeq 0
  }
where we used \truque. The $\simeq$ sign is to remember that the
expression vanishes {\it after} integration over the positions (we
omitted the integral signs to avoid cluttering). Similar
manipulations to the ones of appendix A lead to
  \eqnn\gpu
  \eqnn\gpd
  $$\eqalignno{
  & \qquad \quad \bigl\langle (\lb \g_{mnp} D)\big[ \Omega^1 (\l A^5)(\l\g^m W^2)
  (\l\g^n W^3)(\l\g^p W^4)\bigr] \bigr\rangle \cr
  & \qquad =
  \bigl\langle (\lb \g_{mnp} D)\big[ (\l A^5)\Omega^1 \big](\l\g^m W^2)
  (\l\g^n W^3)(\l\g^p W^4) \bigr\rangle \cr
  & \qquad + 12(\l\lb) \bigl\langle\Omega^1 (\l A^5)\big[ (\l\g^m W^2)
  (\l\g^n W^3){\cal F}^4_{mn} + {\rm cycl (234)} \bigr]\bigr\rangle
  & \gpu\ \cr
  \noalign{\hbox{and}}
  & \qquad \quad \bigl\langle (\lb \g_{mnp} D)\big[ (\l A^5)\Omega^1 \bigr]
         (\l\g^m W^2)(\l\g^n W^3)(\l\g^p W^4) \bigr\rangle \cr
  &\qquad = 4 (\l\lb) \, \bigl\langle \Omega^1 (\l A^5)(\l\g^m W^2)(\l\g^n W^4)
            {\cal F}^3_{mn}\bigr\rangle \cr
  &\qquad + 2 \, \bigl\langle (Q\Omega^1)(\l\g^m W^2)(\l\g^n W^4)\big[(\l A^5)(\lb\g_{mn}W^3)
              - (\l\lb)(A^5\g_{mn}W^3) \bigr] \bigr\rangle \cr
  &\qquad + 2 \, \bigl\langle (\l\lb)(\l A^5)(\l\g^m W^2)(\l\g^n W^4)(W^3\g_{mn}D)\Omega^1 \bigr\rangle \,.
  & \gpd\
  }$$
{}From \totderiv, \gpu\ and \gpd\ it follows that $\d {\cal A} =
\eta(z_2,z_5)\d K_{25} \simeq 0$, where
  \eqnn\gaugefim
  $$\eqalignno{
 (\l\lb) \d K_{25} & =
 + 16(k^2\cdot k^5) \bigl\langle \Omega^1 (\l A^5)
    (\l\g^m W^2)(\l\g^n W^4){\cal F}^3_{mn}\bigr\rangle \cr
  & \quad
  + 12(k^2\cdot k^5)\bigl\langle \Omega^1 (\l A^5)(\l\g^m W^2)
    (\l\g^n W^3){\cal F}^4_{mn} \bigr\rangle \cr
  & \quad
  + 12(k^2\cdot k^5)\bigl\langle \Omega^1 (\l A^5)(\l\g^m W^3)
    (\l\g^n W^4){\cal F}^2_{mn} \bigr\rangle \cr
  & \quad
  + 2(k^2\cdot k^5)\bigl\langle (Q\Omega^1)(\l A^5)(\l\g^m W^2)
    (\l\g^n W^4)(\lb\g_{mn}W^3) \bigr\rangle \cr
  & \quad
  + 2 (k^2\cdot k^5)\bigl\langle (\l A^5)(\l\g^m W^2)(\l\g^n W^4)
    (W^3\g_{mn}D)\Omega^1\bigr\rangle & \gaugefim\ \cr
  & \quad
  - 2 (k^2\cdot k^5)\bigl\langle (Q\Omega^1)(\l\g^m
  W^2)(\l\g^n W^4)(A^5\g_{mn}W^3)\bigr\rangle - (2\leftrightarrow 5).
  }$$
Now it remains to be shown that \gaugefim\ is identical to the
gauge variation of \princ. To see this one uses
$$
\bigl\langle(Q\Omega^1)(\l\g^m W^2)(\l\g^n W^4){\cal
F}^3_{mn}(k^2\cdot A^5)\bigr\rangle
$$
\eqn\gaugea{ = - \bigl\langle \Omega^1(\l\g^m W^2)(\l\g^n
W^4){\cal F}^3_{mn} \big[k^2_r(\l\g^r W^5) + (k^2\cdot k^5)(\l
A^5)\big]\bigr\rangle, } and
$$
 \bigl\langle(Q \Omega^1)(\l\g^m \g^{rs} W^5)(\l\g^n W^4){\cal F}^2_{rs}{\cal F}^3_{mn}\bigr\rangle
- (2\leftrightarrow 5)
$$
\eqn\impo{ = - 4 k^2_r\bigl\langle \Omega^1(\l\g^r W^5)(\l\g^m
W^4) (\l\g^n W^2){\cal F}^3_{mn}\bigr\rangle -(2\leftrightarrow
5), } where the explicit antisymmetry in $[25]$ is important,
otherwise there would be extra terms. Furthermore,
$$
- k^2_m\bigl\langle(Q\Omega^1)(\l\g^{[m} W^3)(\l\g^{n]} W^4)(W^5
\g_n W^2)\bigr\rangle
$$
$$
= - k^2_m\bigl\langle \Omega^1 (\l\g^{[m} W^3)(\l\g^{n]}
W^4)\big[(\l\g^t W^2) {\cal F}^5_{nt} - (\l \g^t W^5){\cal
F}^2_{nt}\big]\bigr\rangle
$$
and finally
$$
- \bigl\langle(Q\Omega^1)(\l\g^{[m} W^3)(\l\g^{n]} W^4){\cal
F}^2_{mt} {\cal F}^5_{nt}\bigr\rangle
$$
\eqn\gaugeb{ = + \bigl\langle \Omega^1 (\l\g^{[m} W^3)(\l\g^{n]}
W^4)\big[ (\l\g^t W^2)k^2_m{\cal F}^5_{nt}  - (\l\g^t
W^5)k^5_m{\cal F}^2_{nt}\big]\bigr\rangle } {}From the above
results one obtains the gauge variation of \princ. It reads
  \eqnn\gaugefimb
  $$\eqalignno{
  (\l\lb) \d K_{25} = & + 16(k^2\cdot k^5) \, \bigl\langle \Omega^1 (\l
  A^5)(\l\g^m W^2)(\l\g^n W^4){\cal F}^3_{mn} \bigr\rangle \cr
  &
  + 12(k^2\cdot k^5) \, \bigl\langle \Omega^1 (\l A^5)(\l\g^m W^2)(\l\g^n
  W^3){\cal F}^4_{mn}\bigr\rangle \cr
  &
  + 12(k^2\cdot k^5) \, \bigl\langle \Omega^1 (\l A^5)(\l\g^m W^3)(\l\g^n
  W^4){\cal F}^2_{mn} \bigr\rangle \cr
  &
  + 2(k^2\cdot k^5) \, \bigl\langle (Q\Omega^1)(\l A^5)(\l\g^m W^2)(\l\g^n
  W^4)(\lb\g_{mn}W^3) \bigr\rangle \cr
  &
  + 2(k^2\cdot k^5) \, \bigl\langle (\l A^5)(\l\g^m W^2)(\l\g^n
  W^4)(W^3\g_{mn}D)\Omega^1\bigr\rangle  & \gaugefimb\ \cr
  &
  - 2(k^2\cdot k^5) \, \bigl\langle (Q\Omega^1)(\l\g^m
  W^2)(\l\g^n W^4)(A^5\g_{mn}W^3)\bigr\rangle - (2\leftrightarrow 5)
  }$$
and it is equal to \gaugefim, as was expected. To get \gaugefimb\
from \princ\ we also used that \eqn\see{
 k^2_r\bigl\langle \Omega^1 (\l\g^r W^5)(\l\g^{m} W^3)(\l\g^n W^4){\cal F}^2_{mn}\bigr\rangle
+ 2 k^2_m\bigl\langle \Omega^1 (\l\g^{[m} W^3)(\l\g^{n]} W^4)(\l
\g^t W^5){\cal F}^2_{nt}\bigr\rangle = 0, } which can be proved by
writing ${\cal F}^2_{nt}=k^2_n A^2_t - k^2_t A^2_n$ and noting the
vanishing of the factor containing $k^2_m k^2_n$ due to the
antisymmetry over $[mn]$. Therefore the second term of \see\ is
equal to
$$
- 2\bigl\langle \Omega^1 (\l\g^{[m} W^3)(\l\g^{n]} W^4)(\l \g^t
W^5)k^2_m k^2_t A^2_n\bigr\rangle = - k^2_t\bigl\langle \Omega^1
(\l\g^t W^5)(\l\g^{m} W^3)(\l\g^n W^4){\cal F}^2_{mn}\bigr\rangle,
$$
which cancels the first.

\newsec{Simplifying the answer}

As one can see from \princ\ the kinematic factor is a bit awkward
and it would be desirable to simplify it using pure spinor
superspace identities. This is accomplished in Appendix B, where
among other things we identified and dropped total derivative
terms in \princ, which simplifies to
  \eqn\pilithia{
  K_{25} = - 40 \Bigl\langle
  (\l A^1) \Big[
  (\l\g^m W^2)(k^2\cdot A^5) - {1\over 4} (\l\g^m\g^{rs} W^5) \cF^2_{rs} \Big]
  (\l\g^n W^3){\cal F}^4_{mn} \Bigr\rangle - (2\leftrightarrow 5).
}

One should note the absence of $\lb_\a$ in the rhs of \pilithia,
which is a remarkable fact in view of \arr. That makes of
\pilithia\ a truly {\it tree-level} pure spinor superspace
integral, which is readily evaluated using the measure
$\langle(\l\g^m\t)(\l\g^n\t)(\l\g^p\t)(\t\g_{mnp}\t)\rangle = 1$
and the method described in the appendix of \anom\ (see also other
methods in \stahn). The gauge transformation of \pilithia\ under
$\d A^1_\a = D_\a \Omega^1$ is simply
\eqn\gaugF{ (\l\lb) \d
K_{25} = 40(k^2\cdot k^5) \langle \Omega^1 (\l A^5) (\l\g^m W^2)
(\l\g^n W^3) {\cal F}^4_{mn}\rangle - (2\leftrightarrow 5),
}
where we used
$(\l\g^m \g^{rs}W^5)k^2_r(\l\g_s W^2) = 2 k^2_r(\l\g^m W^2)(\l\g^r W^5)$ \mafraids.
Furthermore one can observe that \pilithia\ is the result of
computing the $z_5 \rightarrow z_2$ OPE in
  \eqn\legal{
  40 \, \bigl\langle (\l A^1)(\l^0\g^m W^2)(\l^0 \g^n W^3)
  {\cal F}^4_{mn} U^5\bigr\rangle
  }
with the restriction that the pure spinors in $(\l\g^m W^2)$ and
$(\l\g^n W^3)$ have no non-zero modes, but $(\l A^1)$ has. These
$\l^\a$ non zero-modes can be understood by noticing that except
for the pure spinor appearing in the non-integrated vertex --
which comes from the amplitude prescription -- the others are the
zero-modes of $\l^\a$ which appear in the measures \dl\ -- \ds.
Having found \legal, the generalization to the scattering of
closed strings almost suggests itself and could be used to check
the $(t_8 t_8 \pm \e_{10}\e_{10}) R^4 $ sign issue (see e.g.
\richards).

Motivated by the simplicity of \legal, we now investigate whether
this single correlator can reproduce the other contributions to
the amplitude as well. Using \legal\ and the identities of
\mafraids\ one can easily compute the OPEs involving the position
of the unintegrated vertex. For example, the computation as $z_5
\rightarrow z_1$ leads to
\eqn\kuc{
    L_{15} = -40 \, \Bigl\langle
    \Bigl[(\l A^1)(k^1\cdot A^5) + A^1_p (\l\g^p W^5) \Big]
    (\l\g^m W^2)(\l\g^n W^3){\cal F}^4_{mn}
    \Bigr\rangle
}
whose gauge transformation under $\d A^1_\a = D_\a\Omega^1$ can be
easily checked to be
  \eqn\gaugeuc{
  \d L_{15} = + 40 \, (k^1\cdot k^5) \, \bigl\langle
  \Omega^1 (\l A^5)(\l\g^m W^2)(\l\g^n W^3) {\cal F}^4_{mn}
  \bigr\rangle \,. }
With these explicit results for the OPEs involving the position
$z_1$ we can obtain an expression for the full kinematic factor
whose gauge invariance is easily checked. To do this we use that
$$
L_{15}\eta(1,5) = - L_{15}{1\over (k^1\cdot k^5)}\big[ (k^2\cdot
k^5)\eta(2,5) + (k^3\cdot k^5)\eta(3,5) + (k^4\cdot k^5)\eta(4,5)
\big]
$$
(similarly for $L_{1j}$) which follows from the vanishing of $\int
L_{15}(\p/ \p z_5)\exp{\sum (k^i\cdot k^j)f(z_i,z_j)}$ using the
canceled propagator argument. This allows one to express terms
proportional to $\eta(z_1,z_j)$ as a linear combination of
$\eta(z_i,z_j)$, where $i,j\neq 1$. Doing this one obtains
  \eqn\final{
  K_5 = \sum_{2\le i < j \le 5} \Bigl[ K_{ij} -
  L_{1j}{(k^i\cdot k^j)\over(k^1\cdot k^j)} + L_{1i}{(k^i\cdot
  k^j)\over (k^1\cdot k^i)}\Bigr]\eta(z_i,z_j) }
whose gauge
invariance is easily shown by using \gaugF, \gaugeuc\ and the
analogous expressions for the other factors of $K_{ij}$ and
$L_{1j}$. For completeness we also write down the result for the
$L_{15}$ kinematic factor which follows from the
analogous\foot{The actual details are different because it
involves a different set of OPEs to be computed, which also
includes the contribution from $\half N^{mn}{\cal F}_{mn}$.} and
equally long computation leading to \princ,
  \eqnn\Luc
  $$\eqalignno{
  (\l\lb)^2 L_{15} &= -16 (\l\lb)\langle(\l A^1)(\l\g^m W^2)(\l\g^n
  W^4){\cal F}^3_{mn}(k^1\cdot A^5)\rangle \cr
  &\quad
  - 12 (\l\lb)\langle(\l A^1)(\l\g^m W^2)(\l\g^n W^3){\cal
  F}^4_{mn}(k^1\cdot A^5)\rangle \cr
  &\quad
  - 12 (\l\lb)\langle(\l A^1)(\l\g^m W^3)(\l\g^n W^4){\cal
  F}^2_{mn}(k^1\cdot A^5)\rangle \cr
  &\quad
  - 16 (\l\lb)\langle A^1_p (\l\g^m W^2)(\l\g^n W^4){\cal
  F}^3_{mn}(\l\g^p W^5)\rangle \cr
  &\quad
  - 12 (\l\lb)\langle A^1_p (\l\g^m W^2)(\l\g^n W^3){\cal
  F}^4_{mn}(\l\g^p W^5)\rangle \cr
  &\quad
  - 12 (\l\lb)\langle A^1_p (\l\g^m W^3)(\l\g^n W^4){\cal
  F}^2_{mn}(\l\g^p W^5)\rangle \cr
  &\quad
  + 2(k^1\cdot k^5)\langle(\l A^1)(\l A^5)(\l\g^m W^2)(\l\g^n
  W^4)(\lb\g_{mn}W^3)\rangle \cr
  &\quad
  + 2(\l\lb)(k^1\cdot k^5)\langle(\l A^5)(\l\g^m W^2)(\l\g^n
  W^4)(A^1\g_{mn}W^3)\rangle \cr
  &\quad
  - 2(\l\lb)(k^2\cdot k^5)\langle(\l A^1)(\l\g^m W^2)(\l\g^n W^4)(A^5\g_{mn}W^3)\rangle
  \,. & \Luc\
  }$$
We have checked that the component expansion of \final\ does not
change if we use the simplified expressions \pilithia\ and \kuc\
or the longer versions given by \princ\ and \Luc\ (and their
analogous expressions for other labels). That provides a
consistency check that the total derivative terms were correctly
identified.

We conclude that the complete five-point amplitude is given by
\eqn\simple{ {\cal A} = \sum_{\rm top}\int {dt \over t}
\prod_{I=2}^5 \int dz^I K_5(z_1,{\ldots} ,z_5) \langle
\prod_{i=1}^5 {\rm e}^{ik^i\cdot X(z_i,{\bar z}_i)}\rangle } where
-- as alluded to in the introduction -- the kinematic factor is
$$
K_5(z_1,{\ldots},z_5) = \sum_{j=2}^5 L_{1j}\eta(z_1,z_j) + \sum_{2
\le i<j \le 5} K_{ij}\eta(z_i,z_j)
$$
and has the explicit form
  \eqnn\LudKdc
  $$\eqalign{
  L_{12} &= - 40 \, \Bigl\langle
  \Big[(\l A^1)(k^1\cdot A^2) + A^1_p (\l\g^p W^2) \Big]
  (\l\g^m W^5)(\l\g^n W^3){\cal F}^4_{mn}
  \Bigr\rangle \cr
  K_{25} &= - 40 \, \Bigl\langle
  (\l A^1) \Big[
  (\l\g^m W^2)(k^2\cdot A^5) - {1\over 4} (\l\g^m\g^{rs} W^5) \cF^2_{rs} \Big]
  (\l\g^n W^3){\cal F}^4_{mn} \Bigr\rangle - (2\leftrightarrow 5)
  }\LudKdc$$
and analogously for the other labels. Furthermore, as mentioned
above, these kinematic expressions can be obtained from \legal,
which encapsulates all the information about the kinematic
structures appearing at five-points.

\newsec{Bosonic and fermionic component expansions}

\subsec{Matching with RNS, GS and Lee \& Siegel}

The superspace results of the previous section summarize the
computation using the pure spinor formalism. The simplicity of
\LudKdc\ is strong evidence of its correctness, but it must
nevertheless be compared with previous results
\lin\richards\frampton\tsuchiya\ obtained with the RNS and GS
formalisms and the ``ghost pyramid'' approach by Lee and Siegel
(LS) \leeSiegel.

The most straightforward comparison turned out to be with the
five-graviton amplitude calculation by Richards \richards, thanks
to the clarity of the results presented there.  The open-string
parts of that closed-string calculation are clearly identified and
take the form of a kinematic factor $A_{rs}$ (see eq.~(3.20) of
\richards).  Even though the tensors $A_{ij}$ from Richards  and
our expressions for $K_{ij}$ (and similarly $L_{1j}$) are not
exactly the same (they differ by $(k^i\cdot k^j)$ terms) one can
check that the gauge invariant expression \final\ does not change
when $K_{ij}$ is substituted by $A_{ij}$. The conclusion is that
the difference is due to total derivative terms, as they are
automatically eliminated in \final.  The pure spinor formalism is
thus equivalent to the (bosonic) GS result of \richards, and being
a supersymmetric formalism we are guaranteed that the scattering
amplitudes involving fermionic external states are also the
correct ones.

The comparison with the (bosonic) kinematic factor computed by Lee
and Siegel \leeSiegel\ is also done by discarding total derivative
terms. One can easily see the need for that by noticing that their
result features a manifestly gauge invariant integrand (see e.g.
their equation (5.5.1)), whereas
ours is gauge invariant only after integration over the vertex
positions. Therefore one should expect a matching only after total
derivative terms are properly taken into account. Doing that
carefully, we obtain agreement with LS.

More precisely, we first checked that the pole terms of our
expressions agree exactly with the pole terms of equations (5.5.1)
and (5.5.2) in \leeSiegel. This is expected because they are not
affected by total derivative ambiguities.  We compared the
non-pole terms which are proportional to $\eta(z_2,z_5)$, and they
are completely accounted for by adding a total derivative term in
superspace.  To see this explicitly one can add a specially
crafted total derivative term to \pilithia\ as follows:
  $$
  (\l\lb) K_{25} = L_{25} - L_{52}
  $$
  $$
  = -40 \, \bigl\langle (\l A^1)(\l\g^m W^2)(\l\g^n W^4){\cal
  F}^3_{mn}(k^2\cdot A^5)\bigr\rangle + 10 \, \bigl\langle (\l
  A^1)(\l\g^m \g^{rs} W^5)(\l\g^n W^4){\cal F}^2_{rs}{\cal
  F}^3_{mn}\bigr\rangle
  $$
  \eqn\adivinha{ + 20(k^2\cdot k^5) \, \bigl\langle (\l A^5)(\l\g^m
  W^2)(\l\g^n W^4)(A^1\g_{mn} W^3)\bigr\rangle - (2\leftrightarrow
  5). }
The last line is a total derivative in $z_5$: it comes from
  $$
  + 20 \, \bigl\langle (\l\g^m W^2)(\l\g^n W^4)(A^1\g_{mn} W^3)
  \, \p (\l A^5)\bigr\rangle
  $$
as can be seen using $\p (\l A^5) = \p \t^\a \p_\a(\l A^5) + \Pi^m
k^5_m (\l A^5)$.  By explicit computation one can then show that
the five gluon component expansion of \adivinha\ is given by the
simple expression
  \eqnn\both
  $$\eqalignno{
K^{5b}_{25}  = & + {1\over 576} t_8^{mnpqrstu} F^1_{mn}F^2_{pa}F^3_{rs}F^4_{tu}F^5_{qa} \cr
  & + {1\over 1152} t_8^{mnpqrstu}
  \bigl[(k^2\cdot e^5) F^1_{mn}F^2_{pq}F^3_{rs}F^4_{tu}
  -(k^5\cdot e^2)F^1_{mn}F^5_{pq}F^3_{rs}F^4_{tu}\bigr]
  & \both
  }$$
which is the same answer one gets from evaluating the traces in
\leeSiegel.  We therefore conclude the equivalence of the PS and
LS results.

Adopting the ``Schoonschip'' notation by writing objects
contracted into $t_8$ directly on the tensor rather than using
dummy indices, we now rewrite \both\ in terms of polarizations,
  \eqnn\bothwithe
  $$\eqalignno{
  K^{5b}_{25} =
  {1\over 72}\Bigl[ & (k^2\cdot e^5) \, t_8^{e^2(k^2+k^5) e^1 k^1 e^3 k^3 e^4 k^4}
   - (k^5\cdot e^2) \, t_8^{e^5(k^2+k^5) e^1 k^1 e^3 k^3 e^4 k^4} \cr
  &
   - (e^2\cdot e^5) \, t_8^{k^2 k^5 e^1 k^1 e^3 k^3 e^4 k^4}
  - (k^2\cdot k^5) \, t_8^{e^2 e^5 e^1 k^1 e^3 k^3 e^4 k^4} \Bigr] \,.
  & \bothwithe
  }$$
This is, up to
an overall coefficient, the same expression as Richards' $A_{25}$ --
therefore also confirming the equivalence
between the LS and the light-cone GS calculations. It is
reassuring to see that everything matches.

Furthermore one can exploit the superfield nature of the pure
spinor result \adivinha\ to obtain the component expansion
involving fermionic fields. For example, the $b_1f_2b_3b_4f_5$
expansion of \adivinha\ is given by
  \eqn\bothd{
  K_{25}^{2f3b} = {1\over 576}
  t_8^{mnpqrstu} (k^2_m + k^5_m) (\chi^2\g_n \chi^5)
  F^1_{pq} F^3_{rs} F^4_{tu}.
  }
All the other fermionic component expansions are also easily
obtained. However we should note that although \both\ and \bothd\
happen to be gauge invariant, that is not true for other
combination of external particles, and one really needs to use
\final\ to get explicitly gauge invariant results using \adivinha.

\subsec{A different superspace expression for the kinematic factor $K_{ij}$}

It is interesting to note that the very same total derivative term
chosen in \adivinha\ allows us to rewrite $K_{25}$ as\foot{Some
results in this direction were derived in \tese, where it was
shown that the four-point kinematic factor can be rewritten as
$\langle (D\g_{mnp} A) (\l\g^m W)(\l\g^n W)(\l\g^p W) \rangle$.}
  \eqn\DGA{
  5 \, \Bigl\langle (D\g_{mnp}A^1) \Bigl[ (\l\g^m W^2)(k^2\cdot A^5)
  - {1\over 4} (\l\g^m \g^{rs} W^5) {\cal F}^2_{rs} \Bigr]
  (\l\g^n W^3)(\l\g^p W^4) \Bigr\rangle - (2\leftrightarrow 5)
  }
To see this one can use the same kind of manipulations shown in
Appendix B to obtain
$$
\bigl\langle (D\g_{mnp} A^1)(\l\g^m W^2)(\l\g^n W^3)(\l\g^p W^4)
 (k^2\cdot A^5)\bigr\rangle - (2\leftrightarrow 5) =
$$
$$
-8 \, \bigl\langle (\l A^1)(\l\g^m W^2)(\l\g^n W^4){\cal F}^3_{mn}
(k^2\cdot A^5)\bigr\rangle + 4k^2_r\bigl\langle (\l\g^r W^5)\
 (\l\g^m W^2)(\l\g^n W^4)(A^1\g_{mn} W^3)\bigr\rangle
$$
\eqn\expU{
+ 4(k^2\cdot k^5) \, \bigl\langle (\l A^5)(\l\g^m W^2)
(\l\g^n W^4)(A^1\g_{mn}W^3)\bigr\rangle - (2\leftrightarrow 5)
} and
  $$
  \bigl\langle (D\g_{mnp}A^1)(\l\g^m W^3)(\l\g^n W^4)(\l\g^p
  \g^{rs}W^5){\cal F}^2_{rs}\bigr\rangle - (2\leftrightarrow 5) =
  $$
  $$
  = -8 \, \bigl\langle (\l A^1)(\l\g^m\g^{rs} W^5)(\l\g^n W^4){\cal
  F}^2_{rs}{\cal F}^3_{mn} \bigr\rangle
  $$
  \eqn\expD{ + 16k^2_r \, \bigl\langle (\l\g^r W^5)(\l\g^m
  W^2)(\l\g^n W^4)(A^1\g_{mn} W^3) \bigr\rangle - (2\leftrightarrow
  5). }
Combining \expU\ and \expD\ we get
$$
5 \, \bigl\langle (D\g_{mnp}A^1)(\l\g^m W^2)(\l\g^n W^3)(\l\g^p
W^4)(k^2\cdot A^5)\bigr\rangle
$$
$$
-{5\over 4} \, \bigl\langle (D\g_{mnp}A^1)(\l\g^m W^3)(\l\g^n
W^4)(\l\g^p \g^{rs}W^5){\cal F}^2_{rs}\bigr\rangle -
(2\leftrightarrow 5) =
$$
$$
= -40 \, \bigl\langle (\l A^1)(\l\g^m W^2)(\l\g^n W^4){\cal
F}^3_{mn}(k^2\cdot A^5)\bigr\rangle + 10 \, \bigl\langle (\l
A^1)(\l\g^m \g^{rs} W^5)(\l\g^n W^4){\cal F}^2_{rs}{\cal
F}^3_{mn}\bigr\rangle
$$
$$
+ 20(k^2\cdot k^5) \, \bigl\langle (\l A^5)(\l\g^m W^2)(\l\g^n
W^4)(A^1\g_{mn} W^3)\bigr\rangle - (2\leftrightarrow 5),
$$
which is equal to \adivinha, as we wanted to show.

Therefore we have just shown that up to total derivative
terms \DGA\ is an equivalent answer for the massless
five-point kinematic factor at one-loop, which is equal
to the result obtained by Lee \& Siegel in \leeSiegel.

We can also add total derivative terms to $L_{1j}$, for example
$$
L_{12} = - 40\big[\bigl\langle \Big[
(\l A^1) (k^1\cdot A^2) + A^1_p(\l\g^p W^2) \Big]
(\l\g^m W^5)(\l\g^n W^3){\cal F}^4_{mn}\bigr\rangle
$$
\eqn\Lud{ + 20 (k^1\cdot k^2) \, \bigl\langle (\l A^2)(\l\g^m W^5)
(\l\g^n W^3)(A^1\g_{mn}W^4) \bigr\rangle } and one can easily see
that these total derivative terms cancel each other out in \final.
The bosonic components of \Lud\ can again be conveniently written
as
  \eqnn\LudRichS
  $$\eqalignno{
  L^{5b}_{12} =
  {1\over 72}\Bigl[ & (k^1\cdot e^2) \, t_8^{e^1(k^1+k^2) e^3 k^3 e^4 k^4 e^5 k^5}
   - (k^2\cdot e^1) \, t_8^{e^2(k^1+k^2) e^3 k^3 e^4 k^4 e^5 k^5} \cr
  &
   - (e^1\cdot e^2) \, t_8^{k^1 k^2 e^3 k^3 e^4 k^4 e^5 k^5}
  - (k^1\cdot k^2) \, t_8^{e^1e^2e^3k^3e^4k^4e^5k^5} \Bigr] \,,
  & \LudRichS
  }$$
which is the same as Richards' kinematic factor $A_{12}$.

The expression \LudRichS\ also appears in the analysis of the
low-energy limit as equation~(4.14) of \BarreiroHV\ (see also
detailed computations in \ref\MBM{
  R.~Medina, F.~T.~Brandt and F.~R.~Machado,
  ``The open superstring 5-point amplitude revisited,''
  JHEP {\bf 0207}, 071 (2002)
  [arXiv:hep-th/0208121].
}).  There it was shown that $L_{12}/(k^1\cdot k^2)$ is the
``$s$-channel'' part of the two possible field theory diagrams:
when multiplied by $1/(k^1\cdot k^2)$, the first three terms in
\LudRichS\ correspond to a Yang-Mills three-vertex connected by a
propagator to the one-loop\foot{The fact that the kinematic
structure of the tree-level and one-loop amplitudes are both
proportional to $t_8F^4$ makes the tree-level discussion of
formulae (4.14) and (E.1) of \BarreiroHV\ apply equally to our
one-loop case.} $t_8 F^4$ vertex, and the last term is matched by
the non-linear five-point expansion of the one-loop $t_8F^4$
vertex.

The correct factorization of massless poles is also evident in the
two-fermion component expression \bothd\ and can easily be
checked in the bosonic result \both.  The same is true for the
four-fermion terms, where one can for example show that up to
terms proportional to $k^2\cdot k^5$,
  $$
  K_{52}^{4f1b} = - {i \over 144} \Bigl[
  (k^{25}\cdot k^4) (\chi^4 \slashed{e}^1 (\slashed{k}^1 + \slashed{k}^4)
  \slashed{e}^{25} \chi^3)
  + (k^1\cdot k^4) (\chi^4 \slashed{e}^{25} (\slashed{k}^{25} + \slashed{k}^4)
  \slashed{e}^1 \chi^3)  \Bigr]
  $$
which is the form of the four-point one-loop kinematic factor as
displayed in \sen, with $k^{25}=k^2+k^5$ and $e^{25}_a=(\chi^2
\gamma_a \chi^5)$. We have performed this consistency check for
all six inequivalent assignment of zero, two or four fermions to
the five external states.

\lref\stie{
  S.~Stieberger and T.~R.~Taylor,
  ``Non-Abelian Born-Infeld action and type I - heterotic duality. II:
  Nonrenormalization theorems,''
  Nucl.\ Phys.\  B {\bf 648}, 3 (2003)
  [arXiv:hep-th/0209064].
}

The results above suggest that there are no one-loop $F^5$ terms
in the one-loop effective action (as one could expect from the results
of \richards\stie), but we leave the detailed analysis of these
matters for the future.

\vskip 15pt {\bf Acknowledgements:} CRM would like to thank Jos
Vermaseren for quickly fixing (during holidays) a bug in FORM
which was triggered by some pattern matchings used in \pss. He
would also like to thank Kiyoung Lee for having clarified some
points in \leeSiegel, David Richards for his questions answered
about \richards, Stefan Theisen and Marc Magro for discussions,
Ricardo Medina for email exchange and especially Nathan Berkovits
for several discussions, general guidance and suggestions.  CS
thanks Louise Dolan and Ashok Das for discussions, and Jos
Vermaseren for correspondence regarding the trace evaluation in
FORM. CRM acknowledges support by the Deutsch-Israelische
Projektkooperation (DIP H52) (and also FAPESP grant 04/13290-8
during the early stages of this work).  The work of CS was supported by the
U.S.~Department of Energy, grant no. DE-FG-01-06ER06-01, Task~A,
and grant no. DE-FG-02-91ER40685.


\appendix{A}{Covariant evaluation of the four-point kinematic factor}

\noindent
In \anom\ the NMPS kinematic factor of the massless four-point amplitude was shown to be
$$
K_4 = \Bigl\langle (\lb\g_{mnp} D) \bigl[ (\l A^1) (\l \g^m
W^2)(\l \g^n W^3)(\l \g^p W^4) \bigr]\Bigr\rangle
$$
and a proof of equivalence with the minimal pure spinor formalism was presented.
That proof however relied on a U(5) decomposition of variables and ignored the
overall coefficient relating the two versions. A covariant proof \tese\
which overcomes these details will now be presented.

Using the SYM equations of motions and the pure spinor constraint it is
straightforward to show that
$$
 (\lb\g_{mnp} D) \bigl[ (\l A^1) (\l \g^m W^2)(\l \g^n W^3)(\l \g^p W^4) \bigr]
$$
$$
=  \big[(\lb\g_{mnp} D)(\l A^1)\big](\l\g^m W^2)(\l\g^n W^3)(\l\g^p W^4)
$$
\eqn\cool{
+ 12(\l\lb) (\l A^1) \big[(\l \g^m W^4)(\l \g^n W^2){\cal F}^3_{mn}
+  (\l \g^m W^2)(\l \g^n W^3){\cal F}^4_{mn}
+  (\l \g^m W^3)(\l \g^n W^4){\cal F}^2_{mn}\big]
}
Using $\eta_{mn}\g^m_{\a(\b}\g^n_{\g\d)}=0$ and that the factor of
$\big[(\lb\g_{mnp} D)(\l A^1)\big]$ can be substituted by
$\big[(\lb\g_m\g_n\g_p D)(\l A^1)\big]$ due to the pure spinor constraint
we arrive at the following identity
$$
\bigl\langle \big[(\lb\g_{mnp} D)(\l A^1)\big](\l\g^m W^2)(\l\g^n
W^3)(\l\g^p W^4) \bigr\rangle
$$
$$
= \bigl\langle (\lb\g_m\g_n W^3)\Big[(\l\g^n\g_p D)(\l
A^1)\Big](\l\g^m W^2)(\l\g^p W^4)\bigr\rangle
$$
\eqn\legal{ + \bigl\langle (\lb\g_m\g_n \l)(W^3 \g^n\g_p)^{\s}
\Big[D_{\s}(\l A^1)\Big](\l\g^m W^2)(\l\g^p W^4)\bigr\rangle. }
Using $\g^n\g_p = - \g_p\g^n + 2\d^n_p$ and the equation of motion
$Q(\l A)= 0$ the first term of \legal\ vanishes, while the second
becomes
$$
\bigl\langle (\lb\g_m\g_n \l)(W^3 \g^n\g_p)^{\s} \Big[D_{\s}(\l
A^1)\Big](\l\g^m W^2)(\l\g^p W^4)\bigr\rangle
$$
$$
= -2(\l\lb)\bigl\langle (W^3 \g_m\g_p)^{\s}\Big[(\l
D)A^1_{\s}\Big](\l\g^m W^2)(\l\g^p W^4)\bigr\rangle
$$
$$
= + 4(\l\lb)\bigl\langle (\l A^1)(\l\g^m W^2)(\l\g^n W^4){\cal
F}^3_{mn}\bigr\rangle,
$$
where we used $D_\a A_\b + D_\b A_\a = \g^m_{\a\b}A_m$ and integrated
the BRST-charge by parts.
Therefore
$$
K_4 = \Bigl\langle (\lb\g_{mnp} D)\big[(\l A^1)(\l\g^m W^2)(\l\g^n
W^3)(\l\g^p W^4)\big]\Bigr\rangle
$$
$$
= + 16(\l\lb) \, \bigl\langle (\l A^1) (\l \g^m W^4)(\l \g^n
W^2){\cal F}^3_{mn}\bigr\rangle
$$
$$
+ 12(\l\lb) \, \bigl\langle (\l A^1) (\l \g^m W^2)(\l \g^n
W^3){\cal F}^4_{mn}\bigr\rangle + 12(\l\lb) \, \bigl\langle (\l
A^1) (\l \g^m W^3)(\l \g^n W^4){\cal F}^2_{mn}\bigr\rangle \,.
$$
Furthermore, as was shown in \mafraids, all three correlators
above are symmetric in $(234)$, and we obtain \eqn\umloopf{ K_4 =
+40(\l\lb) \, \bigl\langle (\l A^1)(\l\g^m W^2)(\l\g^n W^3){\cal
F}^4_{mn}\bigr\rangle \,, } which is the covariant proof of
equivalence we were looking for.

\appendix{B}{The computation of the kinematic factor $K_{52}$ }

To prove that \arr\ is equal to \princ\ first one uses the SYM equations of
motion and the pure spinor constraint to obtain,
$$
\bigl\langle(\lb\g_{mnp}D)\big[ (\l A^1){\cal F}^2_{rs}
(\l\g^m\g^{rs} W^5)(\l\g^n W^3)(\l\g^p W^4)\big] \bigr\rangle
$$
$$
= \bigl\langle(\lb\g_{mnp}D)\big[ (\l A^1){\cal F}^2_{rs}\big]
(\l\g^m\g^{rs} W^5)(\l\g^n W^3)(\l\g^p W^4)\bigr\rangle
$$
$$
-48(\l\lb)\bigl\langle(\l A^1)(\l\g^{[m}W^3)(\l\g^{n]}W^4){\cal
F}^2_{mu}{\cal F}^5_{nu}\bigr\rangle
$$
$$
+ 12 (\l\lb)\bigl\langle(\l A^1)(\l\g^m \g^{rs}W^5){\cal
F}^2_{rs}\big[ (\l\g^n W^4){\cal F}^3_{mn} + (\l\g^n W^3){\cal
F}^4_{mn}\big]\bigr\rangle
$$
$$
+ 4\bigl\langle(\l\g^{rs}\lb)(\l A^1)(\l\g^m W^3)(\l\g^n
W^4)\big[{\cal F}^2_{rs}{\cal F}^5_{mn} + {\cal F}^5_{rs}{\cal
F}^2_{mn}\big]\bigr\rangle
$$
$$
- 16\bigl\langle(\l\g^{tu}\lb)(\l A^1)(\l\g^{[m} W^3)(\l\g^{n]}
W^4){\cal F}^2_{mt}{\cal F}^5_{nu}\bigr\rangle.
$$
The last two lines vanish after antisymmetrization over $[25]$, therefore we obtain
$$
\bigl\langle(\lb\g_{mnp}D)\big[ (\l A^1){\cal F}^2_{rs}
(\l\g^m\g^{rs} W^5)(\l\g^n W^3)(\l\g^p W^4)\big] \bigr\rangle -
(2\leftrightarrow 5)
$$
$$
=  \bigl\langle(\lb\g_{mnp}D)\big[ (\l A^1){\cal F}^2_{rs}\big]
(\l\g^m\g^{rs} W^5)(\l\g^n W^3)(\l\g^p W^4)\bigr\rangle
$$
$$
+ 12 (\l\lb)\bigl\langle(\l A^1)(\l\g^m \g^{rs}W^5){\cal
F}^2_{rs}\big[ (\l\g^n W^4){\cal F}^3_{mn} + (\l\g^n W^3){\cal
F}^4_{mn}\big]\bigr\rangle
$$
\eqn\parta{ -48(\l\lb)\bigl\langle(\l
A^1)(\l\g^{[m}W^3)(\l\g^{n]}W^4){\cal F}^2_{mu}{\cal
F}^5_{nu}\bigr\rangle
 - (2\leftrightarrow 5).
}
The first line of \parta\ can be rewritten
using $\g^n_{\a(\b}(\g_n)_{\g\d)}=0$, $\{\g^m,\g^n\} = 2\eta^{mn}$ and $(\l\g^m)_\a (\l\g_m)_\b = 0$,
$$
(\lb\g_{mnp}D)\big[ (\l A^1){\cal F}^2_{rs}\big](\l\g^m\g^{rs} W^5)(\l\g^n W^3)(\l\g^p W^4)
$$
$$
= (\l\g_n\g^p D)\big[(\l A^1){\cal F}^2_{rs}\big](\lb\g^m\g^n W^3)(\l\g^m\g^{rs} W^5)(\l\g^p W^4)
$$
$$
+ (\lb\g_m\g^n\l) (W^3\g^n\g^p D)\big[(\l A^1){\cal F}^2_{rs}\big]
(\l\g^m\g^{rs}W^5)(\l\g_p W^4)
$$
$$
= - 8 k^2_r\bigl\langle(\l A^1)(\l\g^m W^2)(\l\g^r W^5)
(\lb\g_{mn} W^3)(\l\g^n W^4)\bigr\rangle
$$
\eqn\partb{
+ 2 (\l\lb)(W^3\g_{mn} D)\big[(\l A^1){\cal F}^2_{rs}\big](\l\g^m\g^{rs}W^5)(\l\g^n W^4),
}
where we used the equation of motion $(\l D)(\l A^1)=0$ and a few gamma matrix identities to
get
$$
2 \bigl\langle(\l D)\big[(\l A^1){\cal F}^2_{rs}\big](\lb\g^m\g^n
W^3) (\l\g^m\g^{rs} W^5)(\l\g^n W^4)\bigr\rangle
$$
$$
= - 8 k^2_r\bigl\langle(\l A^1)(\l\g^m W^2)(\l\g^r W^5)
(\lb\g_{mn} W^3)(\l\g^n W^4)\bigr\rangle
$$
Furthermore, the last term of \partb\ can be rewritten using $D_{\a}(\l A^1) = -(\l D)A^1
+ (\l\g^q)_{\a}A_q$ together with the equations of motion of the SYM superfields,
\eqn\af{
2 (\l\lb)(W^3\g_m\g_p D)\big[(\l A^1){\cal F}^2_{rs}\big](\l\g^m\g^{rs}W^5)(\l\g^p W^4)
}
$$
= - 2 (\l\lb)(W^3\g_m\g_p)^{\a}(Q A^1_{\a}){\cal F}^2_{rs}(\l\g^m\g^{rs}W^5)(\l\g^p W^4)
$$
$$
+ 2(\l\lb)A^1_q(W^3\g^m\g^p\g^q\l){\cal F}^2_{rs}(\l\g^m\g^{rs}W^5)(\l\g^p W^4)
$$
$$
+ 4 k^2_r (\l\lb)(\l A^1)(\l\g^m\g^{rs}W^5)(\l\g^p W^4)
(W^3\g^m\g^p \g^s W^2).
$$
The second line is zero due to the pure spinor condition. Integrating the BRST-charge by
parts \af\ becomes
\eqn\zeu{
 = -{1\over 2}(\l\lb)\bigl\langle (\l\g^{tu}\g^m\g^p A^1)F^3_{tu}F^2_{rs}(\l\g^m\g^{rs}W^5)(\l\g^p W^4)\bigr\rangle
} \eqn\zed{ + {1\over 2}(\l\lb)\bigl\langle (W^3\g_m\g_p
A^1)(\l\g^{mrstu}\l)(\l\g^p W^4){\cal F}^2_{rs}{\cal F}^5_{tu}
\bigr\rangle }
$$
+ 8 k^2_r (\l\lb)\bigl\langle(W^3\g_m\g_p A^1)(\l\g^m W^2)(\l\g^r
W^5)(\l\g^p W^4)\bigr\rangle
$$
$$
+ 4 k^2_r(\l\lb)\bigl\langle (\l A^1)(\l\g^m\g^{rs}W^5)(\l\g^p
W^4) (W^3\g^m\g^p \g^s W^2)\bigr\rangle.
$$
The term \zeu\ is equal to
$
4(\l\lb)(\l A^1)(\l\g^m\g^{rs}W^5)(\l\g^n W^4){\cal F}^2_{rs}{\cal F}^3_{mn},
$
while \zed\
vanishes after antisymmetrization in $[25]$. Therefore, putting all of the above
together we get
$$
\bigl\langle(\lb\g_{mnp}D)\big[ (\l A^1){\cal F}^2_{rs}
(\l\g^m\g^{rs} W^5)(\l\g^n W^3)(\l\g^p W^4)\big] \bigr\rangle -
(2\leftrightarrow 5) =
$$
$$
- 8 k^2_r\bigl\langle(\l A^1)(\l\g^r W^5)(\l\g^m W^2) (\l\g^n
W^4)(\lb\g_{mn} W^3)\bigr\rangle
$$
$$
- 8 k^2_r (\l\lb)\bigl\langle(\l\g^r W^5)(\l\g^m W^2)(\l\g^n
W^4)(A^1\g_{mn} W^3)\bigr\rangle
$$
$$
+ 4 k^2_r(\l\lb)\bigl\langle (\l A^1)(\l\g^m\g^{rs}W^5)(\l\g^n
W^4) (W^2 \g_s \g_{mn} W^3)\bigr\rangle
$$
$$
-48(\l\lb)\bigl\langle(\l A^1)(\l\g^{[m}W^3)(\l\g^{n]}W^4){\cal
F}^2_{mu}{\cal F}^5_{nu}\bigr\rangle
$$
$$
+ 12 (\l\lb)\bigl\langle(\l A^1)(\l\g^m \g^{rs}W^5)(\l\g^n
W^3){\cal F}^2_{rs}{\cal F}^4_{mn}\bigr\rangle
$$
\eqn\fima{ + 16 (\l\lb)\bigl\langle(\l A^1)(\l\g^m
\g^{rs}W^5)(\l\g^n W^4){\cal F}^2_{rs}{\cal F}^3_{mn}\bigr\rangle
 - (2\leftrightarrow 5).
}

{}From \cool\ we can also infer that,
$$
(\lb\g_{mnp}D)\big[ (\l A^1)(k^2\cdot A^5)(\l\g^m W^2)(\l\g^n W^3)(\l\g^p W^4)\big] =
$$
$$
(\lb\g_{mnp}D)\big[ (\l A^1)(k^2\cdot A^5)\big](\l\g^m W^2)(\l\g^n W^3)(\l\g^p W^4)
$$
$$
+ 12(\l\lb) (\l A^1) (k^2\cdot A^5)\big[(\l \g^m W^4)(\l \g^n W^2){\cal F}^3_{mn}
+  (\l \g^m W^2)(\l \g^n W^3){\cal F}^4_{mn}
$$
$$
+  (\l \g^m W^3)(\l \g^n W^4){\cal F}^2_{mn}\big]
$$
and repeating similar steps used to prove \partb\ one gets
$$
(\lb\g_{mnp}D)\big[ (\l A^1)(k^2\cdot A^5)\big](\l\g^m W^2)(\l\g^n W^3)(\l\g^p W^4) =
$$
$$
= - 2k^2_r(\l A^1)(\l\g^r W^5)(\l\g^m W^2)(\l\g^n W^4)(\lb\g_{mn}W^3)
$$
$$
- 2(k^2\cdot k^5)(\l A^1)(\l A^5)(\lb\g_{mn}W^3)(\l\g^m W^2)(\l\g^n W^4)
$$
\eqn\partc{
+ 2 (\l\lb)(W^3\g_{mn} D)\big[(\l A^1)(k^2\cdot A^5)\big](\l\g^m W^2)(\l\g^n W^4)
}
and after several manipulations,
$$
2 (\l\lb)(W^3\g_{mn} D)\big[(\l A^1)(k^2\cdot A^5)\big](\l\g^m W^2)(\l\g^n W^4)
$$
$$
= + 4(\l\lb)(\l A^1)(\l\g^m W^2)(\l\g^n W^4){\cal F}^3_{mn}(k^2\cdot A^5)
$$
$$
+ 2(\l\lb)k^2_r (\l\g^m W^2)(\l\g^n W^4)\big[ (\l A^1)(W^5\g^r \g_{mn} W^3)
- (\l\g^r W^5)(A^1\g_{mn}W^3)\big]
$$
$$
+ 2(\l\lb)(k^2\cdot k^5)(\l\g^m W^2)(\l\g^n W^4)\big[
(\l A^1)(A^5\g_{mn}W^3) - (\l A^5)(A^1\g_{mn}W^3)\big]
$$
And therefore,
$$
(\lb\g_{mnp}D)\big[ (\l A^1)(k^2\cdot A^5)(\l\g^m W^2)(\l\g^n W^3)(\l\g^p W^4)\big]
- (2\leftrightarrow 5) =
$$
$$
= + 2(\l\lb)k^2_r(\l A^1)(\l\g^m W^2)(\l\g^n W^4)(W^5\g^r \g_{mn} W^3)
$$
$$
+ 2k^2_r(\l\g^r W^5)(\l\g^m W^2)(\l\g^n W^4)\big[ (\l A^1)(\lb\g_{mn}W^3)
- (\l\lb)(A^1\g_{mn}W^3)\big]
$$
$$
+ 12(\l\lb) (\l A^1)(k^2\cdot A^5) \big[
  (\l \g^m W^2)(\l \g^n W^3){\cal F}^4_{mn}
+  (\l \g^m W^3)(\l \g^n W^4){\cal F}^2_{mn}\big]
$$
\eqn\fimb{
+ 16(\l\lb)(\l A^1)(\l\g^m W^2)(\l\g^n W^4){\cal F}^3_{mn}(k^2\cdot A^5)
- 2(k^2\cdot k^5) L_{12345} - (2\leftrightarrow 5),
}
where we defined
$$
L_{12345} =  (\l A^1)(\l A^5)(\l\g^m W^2)(\l\g^n W^4)(\lb\g_{mn}W^3)
$$
\eqn\totder{
+ (\l\lb)(\l A^5)(\l\g^m W^2)(\l\g^n W^4)(A^1\g_{mn}W^3)
- (\l\lb)(\l A^1)(\l\g^m W^2)(\l\g^n W^4)(A^5\g_{mn}W^3).
}

Finally from \fima\ and \fimb\ it follows that
$$
K_{52} = {1\over 4}\bigl\langle(\lb\g_{mnp}D)\big[ (\l A^1){\cal
F}^2_{rs} (\l\g^m\g^{rs} W^5)(\l\g^n W^3)(\l\g^p W^4)\big]
\bigr\rangle
$$
$$
- \bigl\langle (\lb\g_{mnp}D)\big[ (\l A^1)(k^2\cdot A^5)(\l\g^m
W^2)(\l\g^n W^3)(\l\g^p W^4)\big]\bigr\rangle
$$
$$
= - 16(\l\lb)(\l A^1)(\l\g^m W^2)(\l\g^n W^4){\cal F}^3_{mn}(k^2\cdot A^5)
$$
$$
- 12(\l\lb) (\l A^1) \big[
  (\l \g^m W^2)(\l \g^n W^3){\cal F}^4_{mn}
+  (\l \g^m W^3)(\l \g^n W^4){\cal F}^2_{mn}\big](k^2\cdot A^5)
$$
$$
+ 3(\l\lb)(\l A^1)(\l\g^m \g^{rs}W^5)(\l\g^n W^3){\cal F}^2_{rs}{\cal F}^4_{mn}
$$
$$
+ 4(\l\lb)(\l A^1)(\l\g^m\g^{rs}W^5)(\l\g^n W^4){\cal F}^2_{rs}{\cal F}^3_{mn},
$$
$$
+ 24(\l\lb) k^2_r (\l A^1)(\l\g^{[n} W^3)(\l\g^{r]} W^4)(W^5 \g^n W^2)
$$
\eqn\grande{
-12 (\l\lb)(\l A^1)(\l\g^{[m} W^3)(\l\g^{n]} W^4){\cal F}^2_{mu}{\cal F}^5_{nu}
+ 2(k^2\cdot k^5) L_{12345} - (2\leftrightarrow 5),
}
where we used $k^2_m(\g^m W^2)_\a = 0$ and
$
k^2_r(\l\g^m \g^r \g^s W^5)(W^2\g^s \g^{mn} W^3) =
$
$$
= - k^2_r (\l\g^m\g^r\g^s
\g^{mn} W^3)(W^5\g^s W^2)
+ k^2_r(\l\g^m \g^r \g^s W^2)(W^5\g^s \g^{mn} W^3)
$$
to get
$$
-k^2_r\bigl\langle (\l A^1)(\l\g^n W^4) \big[ (\l\g^m\g^{rs}W^5)
(W^2 \g_s \g_{mn} W^3) - 2(\l\g^m W^2)(W^5\g^r \g_{mn}
W^3)\big]\bigr\rangle
$$
$$
= 24 k^2_r (\l A^1)(\l\g^{[n} W^3)(\l\g^{r]} W^4)(W^5 \g^n W^2).
$$

Having obtained \grande\ the derivation of \princ\ is now finished, as they are equal.

\subsec{Total derivative terms}

We are going to show that \totder, when multiplied by $(k^2\cdot k^5)$, is part
of a total derivative term which vanishes
when the whole amplitude is integrated over the position of the vertices. To see this
one notes from \partc\ that those terms come from the evaluation of
\eqn\expres{
(\lb\g_{mnp}D)\big[ (\l A^1)(k^2\cdot A^5)\big](\l\g^m W^2)(\l\g^n W^3)(\l\g^p W^4)
}
which is also present in the expressions for $K_{ij}$, where $i,j$ are the same labels
as $(k^i\cdot A^j)$ in \expres.
Therefore -- omitting everything which is not relevant to this proof --
the whole amplitude will contain the following terms,
$$
+  L_{15342} (k^1\cdot k^2) \eta(2,1)
+  L_{12543} (k^1\cdot k^3) \eta(3,1)
+  L_{12354} (k^1\cdot k^4) \eta(4,1)
+  L_{12345} (k^1\cdot k^5) \eta(5,1)
$$
$$
+  (k^2\cdot k^3)\big[ L_{12543}\,  \eta(3,2) +  L_{15342}\, \eta(2,3)\big]
+  (k^3\cdot k^4)\big[ L_{12354}\, \eta(4,3) +  L_{12543}\,  \eta(3,4)\big]
$$
$$
+  (k^2\cdot k^4)\big[ L_{12354}\,  \eta(4,2) +  L_{15342}\, \eta(2,4)\big]
+  (k^4\cdot k^5)\big[ L_{12345}\, \eta(5,4) +  L_{12354}\,  \eta(4,5)\big]
$$
\eqn\all{
+  (k^3\cdot k^5)\big[ L_{12345}\,  \eta(5,3) +  L_{12543}\, \eta(3,5)\big]
+  (k^2\cdot k^5)\big[ L_{12345}\, \eta(5,2) +  L_{15342}\,  \eta(2,5)\big].
}
By pairing the above terms in groups of four one can show that they vanish.
For example,
consider the terms which contain $\eta(5,j)$. It is easy to show that they are
a total derivative in the variable $z_5$,
$$
+  L_{12345} \big[(k^1\cdot k^5) \eta(5,1) +  (k^2\cdot k^5)  \eta(5,2)
+ (k^3\cdot k^5) \eta(5,3) +  (k^4\cdot k^5)  \eta(5,4) =
$$
$$
= -{\p\over \p z_5} \exp \Bigl[ \sum_{i<j}^5 (k^i\cdot
k^j)f(z_i,z_j) \Bigr],
$$
because $\eta(z_5, z_j) = - {\p\over \p z_5}f(z_j, z_5)$. The same can
be shown for the remaining terms of \all,
establishing that they will all vanish by the cancelled propagator argument.

As can be easily inspected, the symmetry in the labels $(34)$ is manifest in \pili\  but
not in \grande. Upon subtraction of the $(3\leftrightarrow 4)$ permutation in \grande\ one
arrives at the following identity
$$
\bigl\langle(\l A^1)\big[4(\l\g^m W^2)(k^2\cdot A^5) - (\l\g^m
\g^{rs} W^5){\cal F}^2_{rs}\big] (\l\g^n W^4){\cal F}^3_{mn}
\bigr\rangle - (2\leftrightarrow 5)
$$
$$
= \bigl\langle(\l A^1)\big[4(\l\g^m W^2)(k^2\cdot A^5) - (\l\g^m
\g^{rs} W^5){\cal F}^2_{rs}\big] (\l\g^n W^3){\cal
F}^4_{mn}\bigr\rangle
$$
$$
+4(k^2\cdot k^5)(\l A^1)(\l A^5)(\l\g^m W^2)(W^3\g_m W^4)
$$
$$
+2(k^2\cdot k^5)(\l A^1)(\l\g^m W^2)\big[(\l\g^n W^3)(A^5\g_{mn} W^4)
- (\l\g^n W^4)(A^5\g_{mn} W^3)\big]
$$
$$
-2(k^2\cdot k^5)(\l A^5)(\l\g^m W^2)\big[(\l\g^n W^3)(A^1\g_{mn} W^4)
- (\l\g^n W^4)(A^1\g_{mn} W^3)\big]
- (2\leftrightarrow 5),
$$
where all terms which contain an explicit $(k^2\cdot k^5)$ factor come from
the total derivative terms as shown in the previous paragraphs\foot{We
nevertheless keep track of them because they are useful when performing component
expansion checks with the computer.}.
The above can be rewritten as
$$
\bigl\langle(\l A^1)\big[4(\l\g^m W^2)(k^2\cdot A^5) - (\l\g^m
\g^{rs} W^5){\cal F}^2_{rs}\big] (\l\g^n W^4){\cal F}^3_{mn}
\bigr\rangle - (2\leftrightarrow 5)
$$
$$
= \bigl\langle(\l A^1)\big[4(\l\g^m W^2)(k^2\cdot A^5) - (\l\g^m
\g^{rs} W^5){\cal F}^2_{rs}\big] (\l\g^n W^3){\cal
F}^4_{mn}\bigr\rangle
$$
\eqn\idtq{
-4(k^2\cdot k^5)(\l A^1)(\l A^5)(\l\g^m W^2)(W^3\g_m W^4) - (2\leftrightarrow 5)
}
where we used $(\l \g^m W^2)\big[(A^1\g^{mn}W^4)(\l\g^n W^3) - (3\leftrightarrow 4)\big]
= 2 (\l\g^m W^2)(\lambda A^1)(W^3\g^m W^4)$.

Furthermore, the relation
$$
Q\big[(W^2\g_{mnp} W^3)(\l\g^{mnp} W^4)
\big] = 24(\l\g^m W^2)(\l\g^n W^3){\cal F}^4_{mn}
$$
$$
-12(\l\g^m W^2)(\l\g^n W^4){\cal F}^3_{mn} - 12(\l\g^m W^4)(\l\g^n W^3){\cal F}^2_{mn}
$$
and the fact that correlations of BRST-trivial operators vanish can be used to show
$$
24\bigl\langle (\l A^1)(\l\g^m W^2)(\l\g^n W^3){\cal F}^4_{mn}\int
U^5\bigr\rangle =
$$
$$
= 12\bigl\langle (\l A^1)(\l\g^m W^3)(\l\g^n W^4){\cal
F}^2_{mn}\int U^5\bigr\rangle + 12\bigl\langle (\l A^1)(\l\g^m
W^2)(\l\g^n W^4){\cal F}^3_{mn}\int U^5\bigr\rangle
$$
\eqn\doiscin{ + \bigl\langle (\l A^1)(W^2 \g_{mnp} W^3)(\l\g^{mnp}
W^4) \int \p (\l A^5)\bigr\rangle. } However care has to be taken
when computing the OPEs in \doiscin\ because of non-commuting
operators. For example, when considering the term appearing in the
derivation of \doiscin, namely $\bigl\langle (\l A^1)Q[(W^2
\g_{mnp} W^3)(\l\g^{mnp} W^4)] \int U^5\bigr\rangle$ one can first
compute the OPEs and then ``integrate'' the BRST charge by parts
or the other way around; and these two operations don't commute.
Using $\p (\l A^5) = \Pi^m k^5_m (\l A^5) + \p\t^\a \p_\a (\l
A^5)$ and with the above caveat one can check that the components
proportional to $\eta(z_2,z_5)$ obey
$$
-24\bigl\langle (\l A^1)(\l\g^m W^3)(\l\g^n
W^4)k^2_{[m}(W^5\g_{n]}W^2)\bigr\rangle -12\bigl\langle (\l
A^1)(\l\g^m W^3)(\l\g^n W^4){\cal F}^2_{mn}(k^2\cdot
A^5)\bigr\rangle
$$
$$
-12\bigl\langle (\l A^1)(\l\g^{[m}W^3)(\l\g^{n]}W^4){\cal
F}^2_{mt}{\cal F}^5_{nt}
$$
$$
= -24\bigl\langle (\l A^1)(\l\g^m W^2)(\l\g^n W^3){\cal
F}^4_{mn}(k^2\cdot A^5)\bigr\rangle + 6\bigl\langle (\l
A^1)(\l\g^m \g^{rs} W^5)(\l\g^n W^3){\cal F}^2_{rs}{\cal
F}^4_{mn}\bigr\rangle
$$
$$
+12\bigl\langle (\l A^1)(\l\g^m W^2)(\l\g^n W^4){\cal
F}^3_{mn}(k^2\cdot A^5)\bigr\rangle - 3\bigl\langle (\l
A^1)(\l\g^m \g^{rs} W^5)(\l\g^n W^4){\cal F}^2_{rs}{\cal
F}^3_{mn}\bigr\rangle
$$
\eqn\coold{ -(k^2\cdot k^5)\bigl\langle (\l A^1)(\l
A^5)(W^2\g_{mnp}W^3)(\l\g^{mnp}W^4)\bigr\rangle -(2\leftrightarrow
5). } where the third term in the left-hand side is related to the
order in which one chooses to compute the OPEs to generate
$\eta(z_2,z_5)$ or to integrate the BRST charge by parts. It is
the result of integrating the BRST charge by parts after computing
the OPE $d_\a(z_5)\t^\b(z_2)$ as $z_5\rightarrow z_2$,
$$
-{1\over 4}(\l A^1)Q\big[ F^2_{rs}(\g^{rs}\g_{mnp}W^3)_\a(\l\g^{mnp}W^4)\big] W_5^\a
- (2\leftrightarrow 5)
$$
$$
= -12\bigl\langle (\l A^1)(\l\g^{[m}W^3)(\l\g^{n]}W^4){\cal
F}^2_{mt}{\cal F}^5_{nt} \eta(5,2) - (2\leftrightarrow 5).
$$
Note also that the term with the
explicit $(k^2\cdot k^5)$ factor also comes from a total derivative
term, and can therefore be dropped in the end\foot{We chose to keep it
to be able to check \coold\ explicitly by a component expansion computation
with FORM \FORM\pss.}.

Substituting \coold\ in \grande\ and then using \idtq\ allows one to obtain
the following expression for $K_{52}$
$$
(\l\lb)^{-1} K_{52} =
$$
$$
-40\bigl\langle (\l A^1)(\l\g^m W^2)(\l\g^n W^3){\cal
F}^4_{mn}(k^2\cdot A^5)\bigr\rangle + 10\bigl\langle (\l
A^1)(\l\g^m \g^{rs} W^5)(\l\g^n W^3){\cal F}^2_{rs}{\cal
F}^4_{mn}\bigr\rangle
$$
$$
+ (k^2\cdot k^5) \big[ 4\bigl\langle (\l A^1)(\l A^5)(\l\g^m
W^2)(W^3\g_m W^4)\bigr\rangle - \bigl\langle (\l A^1)(\l
A^5)(W^2\g_{mnp}W^3)(\l\g^{mnp}W^4)\bigr\rangle\big]
$$
\eqn\ChuckBerry{
+ (k^2\cdot k^5) L_{12345} - (2\leftrightarrow 5),
}
which dropping total derivative terms can be expressed simply as
$$
(\l\lb)^{-1} K_{52} =
$$
$$
-40\bigl\langle (\l A^1)(\l\g^m W^2)(\l\g^n W^3){\cal
F}^4_{mn}(k^2\cdot A^5)\bigr\rangle + 10\bigl\langle (\l
A^1)(\l\g^m \g^{rs} W^5)(\l\g^n W^3){\cal F}^2_{rs}{\cal
F}^4_{mn}\bigr\rangle
$$
\eqn\cumbia{
- (2\leftrightarrow 5),
}
therefore concluding the proof of equation \pilithia.

\appendix{C}{Integrating $\lb_\a$ in pure spinor superspace expressions}

In principle one-loop calculations using the PS formalism will
involve superspace integrals in which the non-minimal pure
spinor $\bar\lambda_\alpha$ does not simply occur as a product $(\lambda\bar\lambda)$, and
one should know how to deal with these expressions. Although in this paper we managed
to isolate this ``new'' type of correlator to the total derivative terms we
will nevertheless show how to compute them explicitly.

As with the $\langle\lambda^3\theta^5\rangle$
correlators, the result is completely fixed by symmetry.  Writing $S^\pm$ for the spinor
irreps of SO(10) and, schematically, $P^\pm$ for pure spinors, we find
the representation content \ref\lie{
    A.~M.~Cohen, M.~van Leeuwen and B.~Lisser,
    LiE v.2.2.2,
    {\tt http://www-math.univ-poitiers.fr/\~{}maavl/LiE/}
}
$$
\eqalign{
\bar\l_\epsilon\l^{(\a}\l^\beta\l^{\gamma}\l^{\delta)} : S^- \otimes \hbox{Sym}^4 S^+ &= 2 \times
[00003] \oplus 1 \times [11010] \oplus \dots \cr
P^- \otimes \hbox{Sym}^4 P^+ &= 1 \times [00003] \oplus \dots \cr
\theta^{[\delta_1} \dots \theta^{\delta_5]}      : {\hbox{Alt}}^5 S^+ &= 1 \times [00030]
\oplus 1 \times [11010] \,,
}
$$
so that there is only one invariant combination of $\bar\l\l^3$ and $\theta^5$.
We will now make use of this uniqueness and construct a spinorial formula relating it
to the $\langle\l^3\theta^5\rangle$ case, where
$$
\bigl\langle \l^\a\l^\beta\l^\gamma \theta^{\delta_1} \cdots \theta^{\delta_5} \bigr\rangle
\equiv \bar{T}^{\a\beta\gamma,\delta_1\dots\delta_5} =
N^{-1} \Bigl[ (\gamma^m)^{\a\delta_1} (\gamma^n)^{\beta\delta_2} (\gamma^p)^{\gamma\delta_3}
(\gamma_{mnp})^{\delta_4\delta_5} \Bigr]_{(\a\beta\gamma)[\delta_1\dots\delta_5]} \,,
$$
with $N$ a normalization constant.  The simplest ansatz is to write
$\langle\bar\l\l^3\theta^5\rangle = \delta\times \bar T$, suitably symmetrized.
However, this time the pure spinor property is essential to the uniqueness argument
and we will need to be careful to subtract gamma traces.  This can be done by applying
the projection operator
\eqn\ptraceless{
{\cal P}^{\a\beta\g\delta}_{\rho\sigma\tau\omega}
  = {1 \over 2772} \Bigl[ \delta^{(\a}_{\rho}\delta^\beta_\sigma\delta^\g_\tau\g\delta^{\delta)}_{\omega}
  - {1\over 4} \delta^{(\a}_{(\rho}\delta^\beta_\sigma\g^a_{\tau\omega)}\g_a^{\g\delta)}
  + {1\over 160} \g^a_{(\rho\sigma} \g_a^{(\a\beta} \g^b_{\tau\omega)} \g_b^{\g\delta)} \Bigr] \,,
}
which is symmetric and gamma-traceless in both sets of indices and satisfies
${\cal P}{\cal P} = {\cal P}$.  We conclude that
$$
\bigl\langle \bar\l_\epsilon \l^{\a_1} \cdots \l^{\a_4} \t^{\delta_1}\cdots \t^{\delta_5} \bigr\rangle
= c\times {\cal P}^{\a_1\dots\a_4}_{\g_1\dots\g_4} \bigl[ \delta^{\g_4}_\epsilon \bar{T}^{\g_1\g_2\g_3,\delta_i} \bigr]
$$
for some constant $c$.  Substituting \ptraceless\ and using that $\bar T$ is gamma-traceless, we get
\eqn\labarformula{
\bigl\langle \bar\l_\epsilon \l^{\a_i} \t^{\delta_i} \bigr\rangle
= {c \over 2772} \Bigl[ \delta^{(\a_1}_\epsilon \bar{T}^{\a_2\a_3\a_4),\delta_i}
- {1 \over 8} \delta^{(\a_1}_{(\g_1} \delta^{\a_2}_{\g_2} (\g^m)_{\g_3)\epsilon} (\g_m)^{\a_3\a_4)} \bar{T}^{\g_i,\delta_i} \Bigr]
}
The normalization is not important for our amplitude calculation, as all our expressions
contain one $\bar\l$ and four $\l$s.  We will set
\eqn\labarnormalize{
\bigl\langle (\bar\l\l) (\l\g^m\t) (\l\g^n\t) (\l\g^p\t) (\t\g_{mnp} \t) \bigr\rangle = 1 \,,
}
which is satisfied for $c=672$.  Equation \labarformula\ can be interpreted in form of a
practical prescription by re-writing it as
\eqn\lbpresc{
\bigl\langle \bar\l_\epsilon \l^{\a_i} \t^{\delta_i} \bigr\rangle
= {2 \over 33} (\l\lb)\Bigl[ \bigl\{ \delta^{\a_1}_\epsilon
\bigl\langle \l^{\a_2} \l^{\a_3} \l^{\a_4} \t^{\d_i} \bigr\rangle
+ \cdots \bigr\} - {1 \over12} \bigl\{ \g_m^{\a_1\a_2}
\bigl\langle \l^{\a_3} \l^{\a_4} (\l\g^m)_\epsilon \t^{\d_i} \bigr\rangle + \cdots \bigr\} \Bigr]
}
where the two curly brackets contain four and twelve terms respectively, corresponding to
the choice of $\a$ indices on $\delta^\a_\epsilon$ and on $\g_m^{\a_i\a_j}$.  We can think of
the first group of terms as coming from eliminating the $\bar\l$ and one of the $\l$s,
contracting their indices.  In the second group, a pair of $\l^{\a_i}\l^{\a_j}$ has
been replaced by $\g_m^{\a_i\a_j}$
and $\lb_\e$ by $(\l\g_m)_\e$.
We have thus reduced everything to the well-known $\langle\l^3\t^5\rangle$ correlators.

For example, applying this procedure to the eighth term in \princ,
we obtain (inside correlators):
  \eqnn\biglb
  $$\eqalignno{
  & (\l A^1) (\l A^5) (\l\g^m W^2) (\l\g^n W^4) (\bar\l\g_{mn} W^3)  = {2\over 33} \Bigl[ \cr
  &
  \Bigl\{ -  (A^1\g_{mn} W^3) (\l A^5) (\l\g^m W^2) (\l\g^n W^4) \
  +  (\l A^1) (A^5\g_{mn} W^3)) (\l\g^m W^2) (\l\g^n W^4)  \cr
  &
  -  (\l A^1) (\l A^5) (W^2 \g^m\g_{mn} W^3)) (\l\g^n W^4)
  +  (\l A^1) (\l A^5) (\l\g^m W^2) (W^4 \g^n \g_{mn} W^3) \Bigr\} \cr
  & - {1\over 12} \Bigl\{\Bigl(
   (A^1\g^a A^5) (\l\g^m W^2) (\l\g^n W^4) - (A^1 \g^a \g^m W^2) (\l A^5) (\l\g^n W^4) \cr
  & \quad\quad + (A^1 \g^a \g^n W^4) (\l A^5) (\l\g^m W^2) +  (\l A^1) (A^5\g^a \g^m W^2) (\l\g^n W^4) & \biglb\ \cr
  & \quad\quad - (\l A^1) (A^5\g^n W^4) (\l\g^m W^2) + (\l A^1) (\l A^5) (W^2 \g^m \g^a W^4) \Bigr) (\l\g_a\g_{mn} W^3) \Bigr\} \Bigr].
  }$$
To check the consistency of \biglb\ one can use the identity
  \eqnn\lbid
  $$\eqalignno{
  &\bigl\langle (\l A^1)(\l A^5) (\l\g^m W^2)\big[(\l\g^n W^4)(\lb\g_{mn} W^3)
  - (3\leftrightarrow 4)\big]\bigr\rangle \cr
  &\qquad = 2 \, \bigl\langle (\l\lb)(\l A^1)(\l A^5)(\l\g^m W^2)(W^3\g_m W^4)
  \bigr\rangle & \lbid\
  }$$
and compute each term of the lhs of \lbid\ using
the corresponding result of \biglb\ and compare it against the
direct computation of the rhs using the standard $\langle\l^3\t^5\rangle$ correlators.
We did this and obtained agreement.

Alternatively,
we can follow the methods of \PSequivII\ and derive tensorial formulae by constructing a symmetry-based
ansatz and using pure spinor identities to relate it to the
normalization condition~\labarnormalize.  Proceeding in this
fashion, we find
  \eqnn\corrlabarBAAAC
  \eqnn\corrlabarBCAAC
  \eqnn\corrlabarBCCAC
  \eqnn\corrlabarDAAAC
  \eqnn\corrlabarDCAAC
  $$
  \eqalignno{
  &\bigl\langle (\bar\l\g^{ab}\l) (\l\g^c\t) (\l\g^d\t) (\l\g^e\t) (\t\g^{rst} \t) \bigr\rangle = {1 \over 140} \Bigl[ \d^{ab}_{cr} \d^{de}_{st} \Bigr]_{[cde][rst]} & \corrlabarBAAAC \cr
  &\bigl\langle (\bar\l\g^{ab}\l) (\l\g^{cde}\t) (\l\g^f\t) (\l\g^g\t) (\t\g^{rst} \t) \bigr\rangle = {1 \over 4620} \Bigl[ \d^{fg}_{ab}\d^{rst}_{cde} + 2 \, \d^{ast}_{cde}\d^g_b\d^f_r - 20 \, \d^{fst}_{cde}\d^g_b\d^a_r \cr
  & \qquad + 23 \, \d^{abt}_{cde}\d^{fg}_{rs} + 24 \, \d^{aft}_{cde}\d^{bg}_{rs} + \d^{fgt}_{cde}\d^{ab}_{rs} \Bigr]_{[ab][cde][fg][rst]} - {1 \over 25200} \epsilon^{abcdefgrst} & \corrlabarBCAAC \cr
  &\bigl\langle (\bar\l\g^{ab}\l) (\l\g^{cde}\t) (\l\g^{fgh}\t) (\l\g^i\t) (\t\g^{rst} \t) \bigr\rangle \cr
  & \qquad = {1 \over 3850} \Bigl[ 36 \, \d^{cde}_{abr}\d^{fgh}_{ist} - 8 \, \d^{cde}_{air}\d^{fgh}_{bst} + 4 \, \d^{cde}_{far}\d^{gh}_{st}\d^i_b - 2 \, \d^{cde}_{fai}\d^{gh}_{st} \d^r_b - 12 \, \d^{cde}_{far}\d^{gh}_{is} \d^t_b \cr
  & \qquad\qquad - 46 \, \d^{cde}_{fab}\d^{ghi}_{rst} - 42 \, \d^{cde}_{fga}\d^{bhi}_{rst} + 2 \, \d^{cde}_{fgi}\d^{abh}_{rst} - 9 \, \d^{cde}_{fgr}\d^{hb}_{st}\d^i_a
  \Bigr]_{[ab][cde][fgh][rst][cde\leftrightarrow fgh]} \cr
  & \qquad + {1 \over 4200} \bigl( \d^{[r|}_{[a} \epsilon_{b]}{}^{cdefghi|st]} + \d^{[r|}_{[c} \epsilon^{ab}{}_{de]}{}^{fghi|st]} - \d^{[r|}_{[f} \epsilon^{abcde}{}_{gh]}{}^{i|st]} \bigr) & \corrlabarBCCAC \cr
  &\bigl\langle (\bar\l\g^{abcd}\l) (\l\g^e\t) (\l\g^f\t) (\l\g^g\t) (\t\g^{rst} \t) \bigr\rangle \cr
  & \qquad = -{1 \over 70} \Bigl[ \delta^{abcd}_{efrs} \delta^{t}_{g} \Bigr]_{[efg][rst]} + {1 \over 25200} \epsilon^{abcdefgrst} & \corrlabarDAAAC \cr
  & \bigl\langle (\bar\l\g^{abcd}\l) (\l\g^{efg}\t) (\l\g^h\t) (\l\g^i\t) (\t\g^{rst} \t) \bigr\rangle
  = {1 \over 1925} \Bigl[ 23 \, \d^{efgr}_{abcd}\d^{st}_{hi} -42 \, \d^{efhr}_{abcd}\d^{st}_{gi} \cr
  & \qquad - 2 \, \d^{efrs}_{abcd}\d^h_g\d^t_i + \d^{ehir}_{abcd}\d^{st}_{fg} + 20 \, \d^{ehrs}_{abcd}\d^{it}_{fg} + \d^{erst}_{abcd}\d^{hi}_{fg} \Bigr]_{[abcd][efg][hi][rst]} \cr
  & \qquad - {1 \over 4200} \bigl( 2 \, \delta^{[a}_{[r}\epsilon^{bcd]efghi}{}_{st]} + \delta^{[e|}_{[r}\epsilon^{abcd|fg]hi}{}_{st]} \bigr) & \corrlabarDCAAC
  }
  $$
This approach is useful if one aims at a direct component
evaluation without being interested in equivalent
$\langle\l^3\t^5\rangle$ correlators, e.g. when checking
superspace manipulations with a computer.

\appendix{D}{Component evaluation of pure spinor superspace correlators}

While amplitude expressions in pure spinor superspace comprise
bosonic and fermionic parts in the form of superfields, it is
often necessary to extract separate components, e.g.~to compare
with existing results.  Comparisons of component expansions also
provide a valuable check on superspace manipulations.  In this
appendix, we summarize some techniques and intermediate results
relevant to the present paper, extending methods published
previously \anom\stahn.

In all but the simplest cases, the evaluation of pure spinor
correlators becomes computationally involved, and the help of a
computer algebra system seems indispensable.  We therefore
emphasize approaches that may be forbidding for a paper-and-pen
calculation, but lend themselves to direct transfer to the
computer.  The present authors have employed independent
implementations of the algorithms, using FORM and Mathematica
respectively.

\subsec{Superfield expansions}

The fermionic expansions of the Yang-Mills superfields satisfy
simple recursion relations, which makes explicit component
expressions readily available \ref\tSYM{
    J.~P.~Harnad and S.~Shnider,
    ``Constraints And Field Equations For Ten-Dimensional Superyang-Mills
    Theory,''
    Commun.\ Math.\ Phys.\  {\bf 106} (1986) 183
\semi
    H.~Ooguri, J.~Rahmfeld, H.~Robins and J.~Tannenhauser,
        ``Holography in superspace,''
        JHEP {\bf 0007}, 045 (2000)
        [arXiv:hep-th/0007104]
\semi
    P.~A.~Grassi and L.~Tamassia,
        ``Vertex operators for closed superstrings,''
        JHEP {\bf 0407}, 071 (2004)
        [arXiv:hep-th/0405072].
}\Rpuri.  In addition, we need the expansion of the antisymmetrised
derivative of the spinor gauge superfield, $D_\a\g_{mnp}^{\a\beta}
A_\beta$.  This object vanishes at order $\t^0$, and the next few
orders can be written as
  \eqn\daexpansion{
  \eqalign{
  &(D\g_{mnp}A)^{(1)} = -{4\over 3}(\t\g_{mnp} \chi) \cr
  &(D\g_{mnp}A)^{(2)} = {1\over 4} (\t \g_{mnp} \slashed{F} \t) + {1\over 4} \partial_r \zeta_s (\t\g^r\g_{mnp}\g^s\t) \cr
  &(D\g_{mnp}A)^{(3)} = (\t\g_{a[mn}\t)(\t\g^a \partial_{p]}\chi) - {1\over 5} (\t\g^{a}{}_{[mn}\t)(\t\g_{p]} \partial_a \chi) \cr
  &(D\g_{mnp}A)^{(4)} = {1\over 96} (\t\g_{[mn}{}^q\t)(\t\g_{p]rs}\t)\partial_q F_{rs} - {7 \over 96} (\t\g_{[mn|t}\t)(\t\g^{rst}\t)\partial_{|p]} F_{rs}\cr
  }}
The first two lines result directly from applying the
supercovariant derivative $D_\a = \partial_\a + {1 \over 2}
(\t\slashed{\partial})_\a$ to the lowest terms in the expansion of
$A_\a$, followed by simple gamma algebra.  For the third line, we
note that the $SO(10)$ representation content of $\t^3$ and of
$\partial_m\chi^\a$ predicts two independent three-forms composed
of these objects.  Indeed, writing $\t^3$ as a $\g$-traceless
spinor-two-form $\Theta^{ab}_\a$ via $\Theta^{ab}_\a = (\t\g^{abc}
\t)(\g_c\t)_\a$, which captures the whole content of $\t^3$
because $(\t\g^{abc}\t)\t^\a = {1\over 2}
(\g^{[a}\Theta^{bc]})^\a$, we can see that those two independent
three-forms are given by $(\Theta_{[mn} \partial_{p]}\chi)$ and
$(\Theta_{a[m}\g_{np]}\partial^a \chi)$.  The easiest way to
obtain their coefficients is then to go to a gamma matrix
representation. Similarly, for the fourth line, there are two
independent three-forms indicated by representation content, and
one can use the Bianchi identity $\partial_{[a}F_{bc]}=0$ as well
as the spinor product identities $(\g_{ab}\t)_\a(\t\g^{abc}\t) =
0$, $(\t\g_a{}^{[mn}\t)(\t\g^{p]qa}\t) = 0$ and
$(\t\g^{mnpqr}{}_{ab}\t)(\t^{sab}\t) = 20
(\t\g^{[mnp}\t)(\t\g^{qr]s}\t)$ to reduce all terms in $(D\g_{mnp}
A)^{(4)}$ to the form given on the right-hand side.

\subsec{Correlator catalog}

The most efficient way to evaluate pure spinor superspace
correlators is to compile a ``catalog'' of building blocks, as
outlined in appendix A of \anom.  This is particularly the case if
one completely automates the process that deconstructs an
arbitrary correlator into these blocks, namely by expanding gamma
products, sorting of spinor bilinears and, for fermionic fields,
applying Fierz rearrangements.  Two different approaches, the
automatic conversion to traces and the component-based evaluation
studied in \stahn, have turned out much slower and have not been
used here\foot{Trace evaluations using Mathematica / GAMMA become
very slow once the number of gamma matrices in the trace reaches
the mid-twenties.  Calculations of traces that take (tens of)
hours with Mathematica typically finish within (tens of) minutes
using FORM.  Still, FORM takes several hours to compute an
expression like \corrlabarBCCAC\ starting from the spinorial
formula \labarformula.  On the other hand, using the ``catalog
method'', the correlator evaluation becomes a matter of seconds. A
difference in performance of about an order of magnitude remains
in favor of FORM, due to the handling of antisymmetric tensors and
of dummy indices in the GAMMA package. }.

In principle the three building blocks $\langle (\l\g^{[5]}\l)
(\l\g^{[1,3,5]}\t) (\t\g^{[3]}\t) (\t\g^{[3]}\t) \rangle$ are
sufficient, but it is more efficient to also allow for the
pairings $\langle (\l\g\t)(\l\g\t)(\l\g\t)(\t\g\t)\rangle$, since
they only need to be computed once, making all subsequent
computations faster.  In addition to the identities listed in
\anom\ and eq. (2.7) of \stahn, we use the following:
  \eqnn\corrftt
  \eqnn\corrffo
  \eqnn\corrfft
  $$\eqalignno{
  &\bigl\langle (\l\g^{a_1\dots a_5}\t)(\l\g^{b_1b_2b_3}\t)(\l\g^{c_1c_2c_3}\t)(\t\g^{d_1d_2d_3}\t) \bigr\rangle = \cr
  &\quad -{3\over 35} \Bigl( \delta^{a_1\dots a_5}_{e_1\dots e_5} + {1 \over 5!} \epsilon^{a_1\dots a_5}{}_{e_1\dots e_5} \Bigr) \Bigl[ ( \delta^{e_1e_2e_3e_4e_5}_{b_1b_2b_3c_1d_1}\delta^{d_2d_3}_{c_2c_3} - \delta^{e_1e_2e_3e_4e_5}_{c_1c_2c_3b_1d_1}\delta^{d_2d_3}_{b_2b_3} ) & \corrftt \cr
  & \qquad + \delta^{e_1e_2e_3e_4e_5}_{b_1b_2c_1c_2d_1}\delta^{d_2d_3}_{b_3c_3} + ( \delta^{e_1e_2e_3e_4e_5}_{b_1b_2c_1d_1d_2}\delta^{b_3d_3}_{c_2c_3} - \delta^{e_1e_2e_3e_4e_5}_{c_1c_2b_1d_1d_2}\delta^{c_3d_3}_{b_2b_3} ) - \delta^{e_1e_2e_3e_4e_5}_{b_1c_1d_1d_2d_3}\delta^{b_2b_3}_{c_2c_3}
   \Bigr]_{[b_i][c_i][d_i]} \cr
  &\bigl\langle (\l\g^{a_1\dots a_5}\t)(\l\g^{b_1\dots b_5}\t)(\l\g^{c}\t)(\t\g^{d_1d_2d_3}\t) \bigr\rangle = -{1\over 7} \Bigl( \delta^{a_1\dots a_5}_{e_1\dots e_5} + {1 \over 5!} \epsilon^{a_1\dots a_5}{}_{e_1\dots e_5} \Bigr) \cr
  & \quad \times \Bigl[ \delta^{e_1e_2e_3e_4e_5}_{b_1b_2b_3b_4d_1}\delta^{d_2}_{b_5}\delta^c_{d_3} - 2 \delta^{e_1e_2e_3}_{b_1b_2b_3} ( \delta^{e_4e_5}_{cd_1}\delta^{d_2d_3}_{b_4b_5} - \delta^{e_4e_5}_{d_2d_3} \delta^{cd_1}_{b_4b_5} )
   \Bigr]_{[b_i][d_i]} & \corrffo \cr
  &\bigl\langle (\l\g^{a_1\dots a_5}\t)(\l\g^{b_1\dots b_5}\t)(\l\g^{c_1c_2c_3}\t)(\t\g^{d_1d_2d_3}\t) \bigr\rangle = {2\over 7} \Bigl( \delta^{a_1\dots a_5}_{e_1\dots e_5} + {1 \over 5!} \epsilon^{a_1\dots a_5}{}_{e_1\dots e_5} \Bigr) \cr
  & \quad \times \Bigl[ \delta^{e_1e_2e_3e_4}_{b_1b_2b_3b_4} ( \delta^{e_5}_{c_1} \delta^{d_1}_{b_5} - \delta^{e_5}_{d_1} \delta^{c_1}_{b_5} ) \delta^{c_2c_3}_{d_2d_3} - 2 \delta^{e_1e_2e_3}_{b_1b_2b_3} ( \delta^{e_4e_5}_{c_1c_2}\delta^{d_1d_2}_{b_4b_5} - \delta^{e_4e_5}_{d_1d_2}\delta^{c_1c_2}_{b_4b_5} ) \delta^{c_3}_{d_3} )
   \Bigr]_{[b_i][c_i][d_i]} & \corrfft
  }
  $$
They have been derived using a symmetry-based ansatz, as described
in \PSequivII.  A useful application of the component-based method
of \stahn\ is that it provides a quick way to compute the
coefficients in such an ansatz.

\subsec{Kinematic reduction}

In the component calculations relevant to this paper, we encounter
kinematic factors that are Lorentz invariant polynomials in the
momenta $k^i_\mu$ as well as the polarization vectors $e^i_\mu$
and/or the fermionic spinor wavefunctions $\chi^{i\a}$.  Due to
the on-shell identities $\sum_i k^i = (k^i)^2 = k^i\cdot e^i =
\slashed{k}^i \chi^i = 0$, there are many relations among these
polynomials, and we need to be able to systematically reduce them
to some sets of independent kinematic invariants.

As will be shown in the following, such reduction algorithms can
be found in the form of a collection of replacement rules, which
can easily be implemented on a computer algebra system.  An
alternative method considered in \stahn\ was based on evaluating
symbolic expressions on a number of integer-valued solutions to
the on-shell identities.  As the number of external
fields and therefore the number of independent kinematic
structures increases, this method becomes unpractical and slow.
However, the component-based method is still useful to ensure that
the end products of the replacement rules are in fact linearly
independent, that is, that no identities have been missed.

For {\it bosonic expressions}, the reduction process is very
simple and consists of eliminating one momentum (say $k^5 \to -k^1
- \cdots -k^4$) in all $k\cdot e$ and $k\cdot k$ products, plus
one additional product like $k^4 \cdot e^5 \to -(k^1+k^2+k^3)\cdot
e^5$, and one $k\cdot k$ combination to resolve the relation ${1
\over 2} (\sum_{i=1}^4 k^i)^2 = \sum_{i<j}^4 k^i\cdot k^j = 0$.
This leaves five independent quadratic momentum invariants, and
three scalar products for every polarization.  Note that any terms
containing the ten-index epsilon tensor $\epsilon_{10}$ will
vanish, since there are only five polarization vectors, and the
five momenta are linearly dependent.

Adding in {\it two fermions} only requires a mild generalization,
as they can be re-written into independent antisymmetric tensors
$(\chi^1 \g^{[k]} \chi^2)$.  Any $\epsilon_{10}$ terms can be
eliminated by dualizing $\epsilon_{10} \g^{[k]} \to \g^{[10-k]}$,
since the $\chi^i$ are chiral spinors.  The only complication
arises from relations due to the Dirac equations $\slashed{k}^1
\chi^1 = \slashed{k}^2 \chi^2 = 0$, but these can be resolved by
replacing
  \eqn\diracreplace{
  (\chi^1 \g^{a_1\dots a_n} \chi^2) k^1_{a_1} \to -(n-1) k^{1[a_2} (\chi^1 \g^{a_3\dots a_n]} \chi^2)
  }
and similarly for $k^2$.  This rule, along with the dualization
step, allows us to regard the spinor bilinears as unconstrained
antisymmetric tensors from now on.  After applying the bosonic
simplification rules and resolving the tensor antisymmetry by
sorting into some order, e.g. $k^2_{a_1}e^5_{a_2}k^1_{a_3}
\g^{a_1a_2a_3\dots} \to e^5_{a_1}k^1_{a_2}k^2_{a_3}
\g^{a_1a_2a_3\dots}$ etc., all remaining structures are
independent.  In the five-point amplitude calculation, where we
are dealing with terms of the form $kkkk\chi^1\chi^2e^3e^4e^5$, we
find 476 independent kinematic structures.

With {\it four fermions}, we have to think about Fierz
transformations which might seem to lead to relations that are
hard to resolve algorithmically.  Fortunately, since we are
dealing with distinct spinors, we can completely avoid this issue
by rearranging all spinor products into the same order, for
example into $(\chi^1 \g^{[k]} \chi^2)(\chi^3 \g^{[l]} \chi^4)$
bilinears, at the outset of the reduction procedure.  We then
eliminate $\g^{[k]}\to \epsilon_{10} \g^{[10-k]}$ whenever $k>5$,
re-write $\epsilon_{10}\epsilon_{10} \to \delta$, dualize
$\epsilon_{10}\g^{[5]} \to \g^{[5]}$, and apply the Dirac equation
replacement \diracreplace\ as well as the bosonic simplification
rules.

The resulting terms will not all be independent, since the duality
properties of the spinor bilinears have not been dealt with, and
this is slightly more involved than in the two-fermion case.  For
example, the contraction of two self-dual five-forms vanishes, so
all terms containing $(\chi^1\g^{a_1\dots
a_5}\chi^2)(\chi^3\g_{a_1\dots a_5}\chi^4)$ have to be set to
zero.  We now show how to resolve this issue for expressions of
the form $\chi^1\chi^2\chi^3\chi^4 e^5 kkk$ as encountered in the
five-point amplitude.  The steps described in the last paragraph
lead to the following structures: First, there are 30 possible
terms in which the spinors combine into a scalar,
  $$
  \Bigl( (\chi^1\g^{a_1}\chi^2)(\chi^3\g_{a_1}\chi^4) \;\hbox{ or }\; (\chi^1\g^{a_1a_2a_3}\chi^2)(\chi^3\g_{a_1a_2a_3}\chi^4) \Bigr) \times (k\cdot k)(e^5\cdot k^{1/2/3}) \,,
  $$
where we have already set to zero the $\g^{[5]}\cdot \g^{[5]}$
products.  Next, there are 116 two-tensor combinations,
  $$
  \eqalign{
  &\Bigl( (\chi^1\g^{m_1}\chi^2)(\chi^3\g^{m_2}\chi^4) \;\hbox{ or }\; (\chi^1\g^{a_1a_2m_1}\chi^2)(\chi^3\g_{a_1a_2}{}^{m_2}\chi^4) \Bigr)
  \cr
  &\times \Bigl( k^{3/4}_{m_1} k^{1/2}_{m_2} (e^5\cdot k^{1/2/3}) \;\hbox{ or }\; k^{3/4}_{m_1} e^5_{m_2} (k\cdot k) \;\hbox{ or }\; e^5_{m_1} k^{1/2}_{m_2} (k\cdot k) \Bigr) \,,
  \cr
  &\Bigl( (\chi^1\g^{a_1m_1m_2}\chi^2)(\chi^3\g_{a_1}\chi^4) \;\hbox{ or }\; (\chi^1\g^{a_1a_2a_3m_1m_2}\chi^2)(\chi^3\g_{a_1a_2a_3}\chi_4) \Bigr)
  \cr
  &\times \Bigl( k^3_{m_1} k^4_{m_2} (e^5\cdot k^{1/2/3}) \;\hbox{ or }\; k^{3/4}_{m_1} e^5_{m_2} (k\cdot k) \Bigr) \,,
  \cr
  &\Bigl( (\chi^1\g^{a_1}\chi^2)(\chi^3\g_{a_1}{}^{m_1m_2}\chi^4) \;\hbox{ or }\; (\chi^1\g^{a_1a_2a_3}\chi^2)(\chi^3\g_{a_1a_2a_3}{}^{m_1m_2}\chi^4) \Bigr)
  \cr
  &\times \Bigl( k^1_{m_1} k^2_{m_2} (e^5\cdot k^{1/2/3}) \;\hbox{ or }\; k^{1/2}_{m_1} e^5_{m_2} (k\cdot k) \Bigr) \,.
  }
  $$
Finally, there are 8 four-tensor combinations,
  $$
  \eqalign{
  &(\chi^1\g^{m_1m_2m_3}\chi^2)(\chi^3\g^{m_4}\chi^4) \times k^3_{m_1} k^4_{m_2} e^5_{m_3} k^{1/2}_{m_4} \,,
  \cr
  &(\chi^1\g^{m_1}\chi^2)(\chi^3\g^{m_2m_3m_4}\chi^4) \times k^{3/4}_{m_1} k^1_{m_2} k^2_{m_3} e^5_{m_4} \,,
  \cr
  &(\chi^1\g^{a_1m_1m_2}\chi^2)(\chi^3\g_{a_1}{}^{m_3m_4}\chi^4) \times \bigl( k^3_{m_1} k^4_{m_2} e^5_{m_3} k^{1/2}_{m_4} \;\hbox{ or }\; k^3_{m_1} k^4_{m_2} e^5_{m_3} k^{1/2}_{m_4} \bigr) \,.
  }
  $$
These 154 structures are independent, as can be seen by going to
components, and it will turn out that they form a complete
kinematic set.  There are three more groups of possible outcomes,
but they can be simplified using duality manipulations.  The first
group contains a four-tensor of the form $\epsilon_{10}
(\chi^1\g^{[3]}\chi^2)(\chi^3\g^{[3]}\chi^4)$ contracted into
$kkke^5$:
  $$
  X_1 = \epsilon^{m_1\dots m_4}{}_{a_1a_2a_3b_1b_2b_3} (\chi^1\g^{a_1a_2a_3}\chi^2)(\chi^3\g^{b_1b_2b_3}\chi^4) \times k^1_{m_1} k^2_{m_2} k^3_{m_3} e^5_{m_4}
  $$
and three others with $k^1k^2k^4$, $k^1k^3k^4$, or $k^2k^3k^4$.
Upon dualising $\epsilon \g^3 \to \g^7$,
  $$
  X_1 = 3! (\chi^1\g_{m_1m_2m_3m_4b_1b_2b_3b_4}\chi^2)(\chi^3\g^{b_1b_2b_3}\chi^4) \times k^1_{m_1} k^2_{m_2} k^3_{m_3} \,,
  $$
the momenta $k^1$ and $k^2$ are contracted into the
$(\chi^1\chi^2)$ bilinear and the Dirac equation can be used to
reduce $\g^{[7]}\to\g^{[5]}$ and thereby relate $X_1$ to the
previous list.  The second group consists of
  $$
  \eqalign{
  &Y_{1/2} = (\chi^1\g^{a_1a_2m_1m_2m_3}\chi^2)(\chi^3\g_{a_1a_2}{}^{m_4}\chi^4) \times k^3_{m_1} k^4_{m_2} e^5_{m_3} k^{1/2}_{m_4} \cr
  &Y_{3/4} = (\chi^1\g^{a_1a_2m_1}\chi^2)(\chi^3\g_{a_1a_2}{}^{m_2m_3m_4}\chi^4) \times k^{3/4}_{m_1} k^1_{m_2} k^2_{m_3} e^5_{m_4}
  }
  $$
and can be reduced similarly.  For example, in the first line,
dualization leads to
  $$
  \Bigl( {1 \over 20} \delta^{d_1d_2m_4}_{m_1m_2m_3} (\chi^1\g^{d_3\dots d_7}\chi^2) - {1 \over 12} \delta^{d_1d_2d_3}_{m_1m_2m_3} (\chi^1\g^{d_4\dots d_7m_4}\chi^2) \Bigr) (\chi^3\g_{d_1\dots d_7}\chi^4) \times k^3_{m_1} k^4_{m_2} e^5_{m_3} k^{1/2}_{m_4} \,,
  $$
and now at least one momentum is contracted with a bilinear where
the Dirac equation can be used.  The third group contains two
$\g^{[5]}$ factors, either as two-tensor
  $$
  Z^{m_1m_2} = (\chi^1\g^{a_1a_2a_3a_4m_1}\chi^2)(\chi^3\g_{a_1a_2a_3a_4}{}^{m_2}\chi^4)
  $$
contracted into at least one momentum, or as four-tensor
  $$
  Z^{m_1m_2,m_3m_4} = (\chi^1\g^{a_1a_2a_3m_1m_2}\chi^2)(\chi^3\g_{a_1a_2a_3}{}^{m_3m_4}\chi^4)
  $$
contracted into three momenta and the polarization $e^5$.  Here we
note that the only two-tensor irrep contained in the tensor
product of two self-dual five-forms is the symmetric traceless
one, so we must have $Z^{[m_1m_2]}=0$.  Similarly, the product
does not contain a completely antisymmetric four-tensor, and hence
$Z^{[m_1m_2,m_3m_4]} = 0$, which implies
  $$
  Z^{m_1m_2,m_3m_4} = Z^{m_1m_3,m_2m_4} - Z^{m_1m_4,m_2m_3} - Z^{m_2m_3,m_1m_4} + Z^{m_2m_4,m_1m_3} - Z^{m_3m_4,m_1m_2} \,.
  $$
In all cases, the symmetries of $Z$ allow us to shuffle at least
one momentum onto a bilinear where the Dirac equation can be used,
relating all terms containing $Z$ tensors to the list of 154
independent structures.  This concludes the simplification of
four-fermion terms.

\listrefs

\end